\documentclass[11pt]{article}
\usepackage{amsmath,amssymb}
\usepackage{graphicx,color}

\numberwithin{equation}{section}

\def\({\left(}
\def\){\right)}

\newcommand{\de}{\partial}
\newcommand{\be}{\begin{equation}}
\newcommand{\ba}{\begin{eqnarray}}
\newcommand{\ea}{\end{eqnarray}}
\newcommand{\ee}{\end{equation}}

\newcommand{\ca}{\mathcal}

\newcommand{\f}{\frac}
\newcommand{\s}{\sqrt}

\newcommand{\ti}{\tilde}
\newcommand{\ap}{\alpha}

\newcommand{\ddd}{\cdot\cdot\cdot}
\newcommand{\no}{\nonumber \\}
\newcommand{\la}{\langle}
\newcommand{\lb}{\rangle}
\newcommand{\ep}{\epsilon}

\newcommand{\ov}{\overline}
 
 \def\de{\partial}

 \def\f {\frac}
 \def\ti{\tilde}
 \def\ap{\alpha}

 \def\ddd{\cdot\cdot\cdot}
 \def\no{\nonumber \\}

 \def\la{\langle}
 \def\lb{\rangle}
 \def\ep{\epsilon}

\begin{document}

\begin{titlepage}
\thispagestyle{empty}

\begin{flushright}
YITP-13-14\\
IPMU13-0045\\
\end{flushright}

\vspace{.4cm}
\begin{center}
\noindent{\Large \textbf{Holographic Local Quenches and Entanglement Density}}\\
\vspace{2cm}

Masahiro Nozaki $^{a}$,
Tokiro Numasawa $^{a}$,
and Tadashi Takayanagi $^{a,b}$

\vspace{1cm}
  {\it
 $^{a}$Yukawa Institute for Theoretical Physics,
Kyoto University, \\
Kitashirakawa Oiwakecho, Sakyo-ku, Kyoto 606-8502, Japan\\
\vspace{0.2cm}
 $^{b}$Kavli Institute for the Physics and Mathematics of the Universe,\\
University of Tokyo, Kashiwa, Chiba 277-8582, Japan\\
 }

\vskip 2em
\end{center}

\vspace{.5cm}
\begin{abstract}
We propose a free falling particle in an AdS space as a holographic
model of local quench. Local quenches are triggered by local
excitations in a given quantum system. We calculate the
time-evolution of holographic entanglement entropy. We confirm a
logarithmic time-evolution, which is known to be typical in two
dimensional local quenches. To study the structure of quantum
entanglement in general quantum systems, we introduce a new quantity
which we call entanglement density and apply this analysis to
quantum quenches. We show that this quantity is directly related to
the energy density in a small size limit. Moreover, we find a simple
relationship between the amount of quantum information possessed by
a massive object and its total energy based on the AdS/CFT.
\end{abstract}

\end{titlepage}

\newpage

\begin{scriptsize}
\tableofcontents
\end{scriptsize}

\newpage

\section{Introduction}

Quantum entanglement provides us with a powerful method of
investigating various quantum states and classifying their
quantum structures. Moreover, quantum entanglement is a useful
tool when we would like to study excited quantum systems which
are far from thermal equilibrium, for example, those under
thermalization processes. Even though in such systems we cannot
define the thermodynamical entropy and temperature etc, we can
always define the entanglement entropy (refer to e.g. the
reviews \cite{Ereview,CCreview,CHreview,Lreview,HEEreview}).

 Quantum quenches provide us with ideal setups to study thermalizations of
quantum systems, which can be realized even in real experiments such as cold atoms.
They are unitary evolutions of pure states triggered by sudden change of parameters such as mass gaps or coupling constants \cite{CaCaG,Eisler,CaCaL}. By using the AdS/CFT \cite{Maldacena,GKPW,AdSreview}, we can relate this problem to the dynamics of gravitational theories \cite{QuenchHol,QuenchHEE,TaUg,Roberts:2012aq}.

One typical class of quantum quenches is called global quenches and
they occur from homogeneous changes of parameters \cite{CaCaG}. Holographic duals of global quenches are dual to black hole formations as have been discussed in \cite{QuenchHol,TaUg}.
 A useful quantity to understand how thermalizations occur is the entanglement entropy \cite{CaCaG}. The holographic entanglement entropy (HEE) \cite{RT,HRT,HEEreview} has been calculated for various quantum quenches \cite{QuenchHEE,Bla,TSSA}. Refer also to \cite{NTT}
for a computation of HEE for a stationary system which is described by an excited pure state dual to an AdS plane wave. In general, non-local probes such as the entanglement entropy are useful to measure the thermalization time. On the other hand, local quantities such as an expectation value of
energy momentum tensor shows an immediate thermalization and is not suitable to see if a given system is completely thermalized \cite{Bla}.
In this sense, the entanglement entropy can serve as a non-equilibrium substitute of thermodynamical entropy.

Another class of quantum quenches is the local quench and this is
triggered by a shift of parameters within a localized region or simply at a point.
One of the aims of the present paper is to provide a simple construction of
holographic dual for local quench and is to calculate
the holographic entanglement entropy. A local quench shows
how localized excitations in a given quantum system propagate to other
spatial regions. Local quenches have been studied
in two dimensional CFTs \cite{Eisler,CaCaL}.
However, local quenches in higher dimensions
have not been understood well. This partially motivates
us to study the local quenches in AdS/CFT, which often
allows us higher dimensional calculations.

We will argue that a simple holographic description of a system just after the
local quench is a free falling particle-like object in an AdS space. It is pulled
into the horizon of AdS space due to the gravitational
force and this is the reason why we observe the non-trivial
time-dependence of entanglement entropy. Therefore this
problem is deeply connected to a fundamental question:
what is the non-gravitational (or CFT) counterpart of gravitational force
via holography ? We will suggest an intuitive answer to this question in the end.

The time evolution of quantum entanglement under local quenches is more
complicated than the global quenches because it is inhomogeneous. To understand its structure clearly we introduce a new quantity which we call entanglement density. It is defined by taking the derivatives of the entanglement entropy $S_A$ with respect to the positions of two boundary points of the subsystem $A$. This quantity counts the number of entangled pairs at a given position. The strong subadditivity guarantees that this quantity is always positive. As we will see this analysis reveals the detailed structure of quantum entanglement under local quenches as well as global quenches.

One more motivation to study local quenches is to estimate the
amount of quantum information possessed by a massive object or
radiations. We will employ the entanglement entropy for local
quenches in order to measure the amount of information included in a
localized excited lump. We will evaluate this quantity by using the
holographic entanglement entropy (HEE) and obtain the simple
conclusion that it is given by the total energy of the object times
its size up to a numerical factor.

The paper is organized as follows: In section two, we explain our
holographic setup of local quench using a free falling particle in
AdS$_{d+1}$. We calculate the holographic energy stress tensors in
this model. In section three, we compute the holographic
entanglement entropy for $d=2,3,4$ assuming that the back-reaction
due to the falling particle is very small. In section four, we
perform an exact analysis of holographic entanglement entropy for
$d=2$. In section 5, we introduce a new quantity which we call
entanglement density and we investigate the evolution of quantum
entanglement structures under local quenches by using this.
In section 6, we study the relation between the amount of information of
an object and its total energy using the results in the previous sections. We
also interpret our results of local quenches using the
idea of the entanglement renormalization and discuss the holographic
interpretation of gravitational force.
 In section 7 we
summarize our conclusions and discuss future problems. In appendix A, we show an explicit
perturbative calculation of back-reactions due to the falling
particle.

\section{Holographic Local Quenches as Falling Particles}

\begin{figure}[ttt]
   \begin{center}
      \includegraphics[height=5cm]{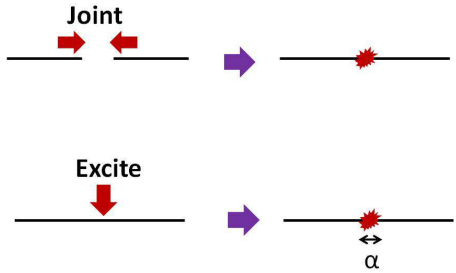}
   \end{center}
   \caption{Setups of local quenches. The upper picture describes
   a process of jointing two systems which are defined on semi-infinite
   lines. The lower one describes localized excitations on an
   infinitely extended system. We define the parameter $\ap$ which measures the size of
   excited region at the beginning of the local quench.
   }\label{fig:quench}
\end{figure}

Local quenches in quantum systems are triggered by a sudden local
change of the Hamiltonian at a specific time. One typical class of
examples will be joining two separated semi-infinite systems at each
endpoint as studied in \cite{Eisler,CaCaL} (see the upper picture in
Fig.\ref{fig:quench}). When this quench process happens, an
interaction between two endpoints is instantaneously introduced.
From the viewpoint of the new Hamiltonian, an locally excited state
is generated just after this local quench. Therefore we can
generally characterize a local quench by local excitations. These
excitations will propagate to other regions under the
time-evolution.

\begin{figure}[ttt]
   \begin{center}
      \includegraphics[height=6cm]{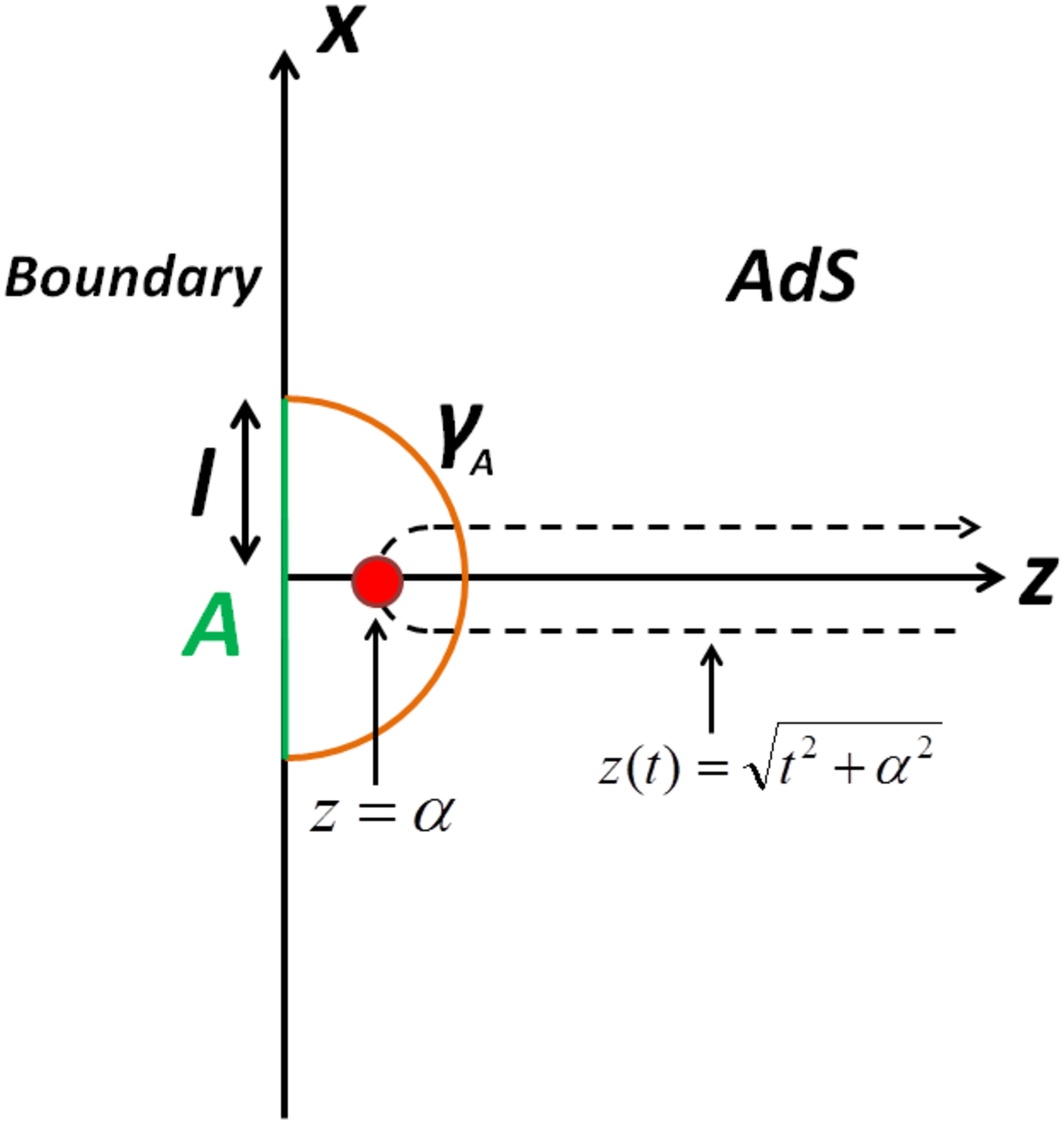}
      \includegraphics[height=6cm]{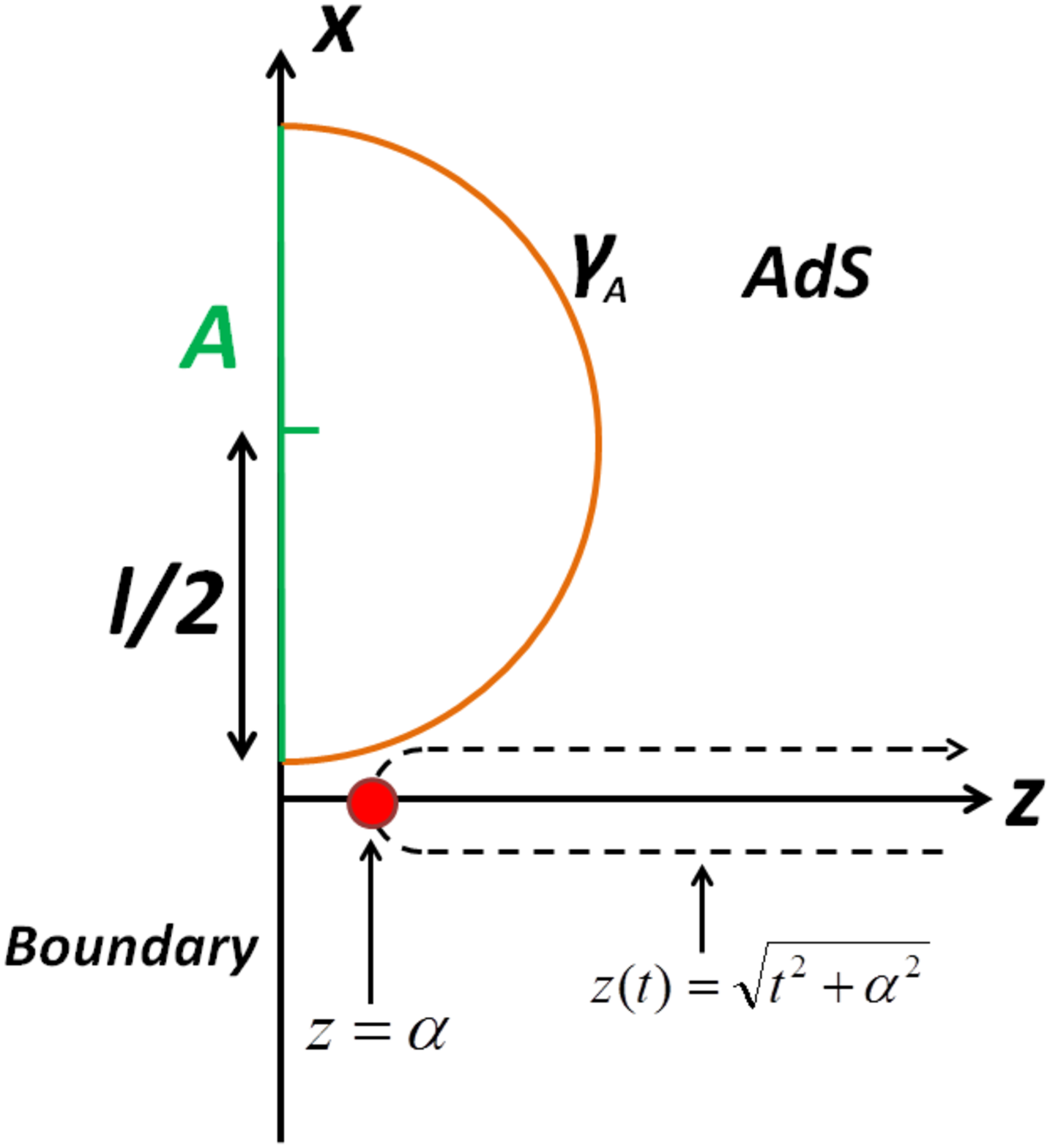}
   \end{center}
   \caption{A falling massive particle in AdS and the calculation
   of holographic entanglement
   entropy for two different choices of the subsystem $A$. It is clear from this picture that the back reaction due to the falling particle gets significant when $l=z(t)$ in the left picture and $t=0$ in the right one because the particle is on top of $\gamma_A$.
    }\label{fig:setup}
\end{figure}

Now we would like to construct gravity duals for local quenches via
the AdS/CFT. Even though we have not found a simple gravity dual of
the original model i.e. joining two CFTs, it is not difficult to
find a holographic model for local excitations. It is given by a
falling massive particle in a Poincare AdS space (see
Fig.\ref{fig:setup}). At $t=0$, the particle is situated near the
AdS boundary and its back-reaction to the metric is highly localized
near the particle. Under its time evolution it falls into the AdS
horizon and its back-reaction spreads out. In the dual CFT, at
$t=0$, the excitations are concentrated in a small localized region
(see the lower picture in Fig.\ref{fig:quench}), whose radius is
defined to be $\ap$. Therefore we can regard this state at $t=0$ as
the one just after the local quench. Later the excitations expand at
the speed of light as we will see e.g. from its holographic energy
stress tensor later. In this way, we can regard this setup as a
gravity dual of local quench.

We will employ the beautiful construction of back-reacted solutions
found by Horowitz and Itzhaki in \cite{HoIt}. The basic idea is to
start from a black hole in a global AdS space and map it into a
Poincare AdS by the coordinate transformation. This leads to a
falling black hole solution. Though we are thinking of a massive
particle with a finite size or equally a star, instead of a black
hole, the asymptotic solution which is outside of the star is the
same as that for a black hole as usual.

\subsection{A Falling Massive Particle in AdS}

Consider a $d+1$ dimensional AdS space (AdS$_{d+1}$) in the Poincare coordinate
\be
ds^2=R^2\left(\f{dz^2-dt^2+\sum_{i=1}^{d-1}dx_i^2}{z^2}\right). \label{ads}
\ee
The radius of AdS is defined to be $R$ and
the coordinate of AdS is represented by $x^\mu=(z,t,x_1,\ddd,x_{d-1})$.

In this AdS space, we introduce a massive object (mass $m$) with a
very small size which is larger than the Schwartzschild radius. Its
motion in the AdS space is described by the trajectory
$x^\mu=X^\mu(\tau)$. In general, the action of a particle with mass
$m$ in a spacetime defined by the metric $g_{\mu\nu}$ is given by
\be S_{p}=-m\int d\tau \int
dx^{d+1}\delta^{(d+1)}(x^\mu-X^\mu(\tau))\s{-g_{\mu\nu}(x)\cdot
\de_\tau  X^\mu(\tau)\cdot \de_\tau X^\nu(\tau)}. \ee

We assume that the particle is situated at $X^i=0$ and we gauge fix by setting $X^t(\tau)=\tau$.
Then the trajectory is specified by the function $X^z(\tau)=z(\tau)$. In the pure AdS
background (\ref{ads}), the action looks like
\be
S=-mR\int dt \f{\s{1-\dot{z}(t)^2}}{z(t)}. \label{eomp}
\ee

The solution to the equation of motion derived from (\ref{eomp}) is given by
\be
z(t)=\s{(t-t_0)^2+\ap^2}, \label{traj}
\ee
where $t_0$ and $\ap$ are integration constants. Below we will set $t_0=0$ by using the time translation invariance. When $t<0$ the particle moves from the horizon to the boundary.
It reaches $z=\ap$ at $t=0$. Later ($t>0$), it again falls into the horizon as
depicted in Fig.\ref{fig:setup}. Thus the
energy of the particle in the AdS space is calculated as
\be
E=\f{mR}{\ap}. \label{energy}
\ee

\subsection{Einstein Equation}

The gravity action coupled to the massive particle reads
\be
S_{tot}=\f{1}{16\pi G_N}\int dx^{d+1} \s{-g}({\ca R}-2\Lambda)+S_{p}\ ,
\ee
where the cosmological constant is given by $\Lambda=-\f{d(d-1)}{2R^2}$ and $G_N$ is the Newton constant.

The equation of motion becomes
\be
{\ca R}^{\mu\nu}-\f{1}{2}g^{\mu\nu}{\ca R}+\Lambda g^{\mu\nu}={\ca T}^{\mu\nu}, \label{EMM}
\ee
where the bulk energy-stress tensor ${\ca T}^{\mu\nu}$ is given by
\be
{\ca T}^{\mu\nu}=\f{8\pi mG_N}{\s{-g}}\cdot \f{\de_tX^\mu\de_tX^\nu}
{\s{-g_{\mu\nu}\cdot \de_t  X^\mu(t)\cdot \de_t X^\nu(t)}}\cdot \delta(z-z(t))\cdot \delta^{d-1}(x_i).
\label{emtsnr}
\ee

We will show a direct perturbative calculations of this back-reaction in appendix A.
 However, below we will take a different step in order to analytically construct the back-reacted solutions.  See the paper \cite{DKK} for analytical calculations of back-reactions to a scalar field in an AdS space. Refer also to \cite{FGMP} for a more extensive analysis and a relation to expanding qluon plasmas, where the back-reacted solutions are called conformal solitons (see \cite{Ran} for spacetime structures of
 conformal solitons).

\subsection{Coordinate Transformation from Global AdS to Poincare AdS}

Now consider the global AdS$_{d+1}$ space defined by the metric
\be
ds^2=-(R^2+r^2)d\tau^2+\f{R^2dr^2}{R^2+r^2}+r^2d\Omega_{d-1}^2. \label{global}
\ee

We can show this is (locally) equivalent to the Poincare AdS$_{d+1}$
space (\ref{ads}) via the following coordinate transformation: \ba
&&
\s{R^2+r^2}\cos\tau=\f{R^2e^{\beta}+e^{-\beta}(z^2+x^2-t^2)}{2z},\no
&& \s{R^2+r^2}\sin\tau=\f{Rt}{z},\no && r\Omega_{i}=\f{Rx_i}{z} \ \
\ (i=1,2,\ddd,d-1),\no &&
r\Omega_d=\f{-R^2e^{\beta}+e^{-\beta}(z^2+x^2-t^2)}{2z}.
\label{corf} \ea Here, the coordinate of $S^{d-1}$ is described by
$(\Omega_1,\Omega_2,\ddd,\Omega_{d})$ such that $\sum_{i=1}^d
(\Omega_i)^2=1$. Also we defined $x^2=\sum_{i=1}^{d-1}x_i^2$. The
arbitrary constant $\beta$ is introduced for the later purpose,
which corresponds to the boost transformation of $SO(2,d)$ symmetry.
If we set $\beta=0$, (\ref{corf}) is reduced to the standard one
which can be found in e.g. \cite{AdSreview}.

\subsection{Back-reacted Metric for a Falling Massive Particle}

In the global coordinate, we can consider a static particle situated at $r=0$.
Following the idea in \cite{HoIt}, we would like to map it into the Poincare AdS.
After the coordinate transformation (\ref{corf}), its trajectory is mapped into
\be
x_i=0, \ \ \ \ z^2-t^2=R^2e^{2\beta}.
\ee
Thus this corresponds to the previous trajectory (\ref{traj}) with the identification
\be
\ap=Re^{\beta}. \label{aprel}
\ee

The back-reacted geometry outside of the massive object is obtained
from the AdS black hole solution \cite{Witten}: \be
ds^2=-\left(r^2+R^2-\f{M}{r^{d-2}}\right)d\tau^2+\f{R^2dr^2}{R^2+r^2-M/r^{d-2}}+r^2d\Omega_{d-1}^2.
\label{sbh} \ee Note that in the AdS$_3$ case ($d=2$), the solution
(\ref{sbh}) for $M<R^2$ is not a black hole solution but a solution
with a deficit angle.\footnote{We do not have to worry about the
singularity because we replace the region near it with a star
solution.} The mass parameter $M$ in (\ref{sbh}) is related to the
mass $m$ of the particle via \be
m=\f{(d-1)\pi^{d/2-1}}{8\Gamma(d/2)}\cdot \f{M}{G_N R^2}.
\label{mrel} \ee

Therefore, we can find the back-reacted metric by performing the coordinate transformation
(\ref{corf}) to the metric (\ref{sbh}). This can be done in a straightforward manner by noting
\ba
&& r=\f{1}{2z}\s{R^4e^{2\beta}+e^{-2\beta}(z^2+x_i^2-t^2)^2-2R^2(z^2-x^2-t^2)},\no
&& d\tau^2=d(\cos\tau)^2+d(\sin\tau)^2,\ \ \ d\Omega_{d-1}^2=\sum_{i=1}^{d}(d\Omega_{i})^2.
\ea

\subsection{Holographic Energy Stress Tensor}

One way to understand the time evolution of the CFT$_{d}$ state dual
to the falling particle in AdS$_{d+1}$, is to calculate the
holographic energy stress tensor. For this purpose, it is useful to
employ the Fefferman-Graham gauge of the coordinates given by the
expression \ba && ds^2=R^2\cdot \f{dz^2+g_{ab}(x,z)dx^adx^b}{z^2}.
\label{FGc} \ea where $x^a=(t,x_i)$. We are considering the case
where the boundary metric $g_{ab}(x,0)$ coincides with the flat
Minkowski metric $\eta_{ab}$ as in the Poincare AdS. Then near the
AdS boundary $z=0$, $g_{ab}(x,z)$ behaves like \be
g_{ab}(x,z)=\eta_{ab}+t_{ab} z^{d}+O(z^{d+1}). \label{emh} \ee In
this setup, the holographic energy stress tensor \cite{EMtensor} is
calculated from the formula: \be T_{ab}=\f{d\cdot R^{d-1}}{16\pi
G_N}\cdot t_{ab}. \label{emht} \ee

Note that the metric we find from the coordinate transformation
(\ref{corf}) of (\ref{sbh}) is not in the form of the
Fefferman-Graham gauge (\ref{FGc}). Thus we need to perform a
coordinate transformation further to achieve this gauge so that we
can employ (\ref{emht}). It is also useful to define the light-cone
coordinate $u=t-\rho$ and $v=t+\rho$, where $\rho=x_1$ for $d=2$ and
$\rho=\s{\sum_{i=1}^{d-1} x_i^2}$ for $d>2$. Finally we obtain the
following energy stress tensor $T_{ab}$ for $d=2,3,4$: \ba && d=2\
(\mbox{AdS}_3): \no
&& T_{uu}=\f{M\ap^2}{8\pi G_N R(u^2+\ap^2)^2}, \ \ \ T_{vv}=\f{M\ap^2}{8\pi G_N R(v^2+\ap^2)^2},\ \ \ T_{uv}=0. \\
&& \no
&& d=3\  (\mbox{AdS}_4):  \no
&& T_{uu}=\f{3M\ap^3}{8\pi G_N R(u^2+\ap^2)^{\f{5}{2}}\s{v^2+\ap^2}},\ \ T_{uv}=\f{M\ap^3}{8\pi G_N R(u^2+\ap^2)^{\f{3}{2}}(v^2+\ap^2)^{\f{3}{2}}},\no
&& T_{vv}=\f{3M\ap^3}{8\pi G_N R(v^2+\ap^2)^{\f{5}{2}}\s{u^2+\ap^2}},\ \ T_{\theta\theta}=\f{3M\ap^3(u-v)^2}{8\pi G_N R(u^2+\ap^2)^{\f{3}{2}}(v^2+\ap^2)^{\f{3}{2}}},
\label{EMthree} \\
&& \no
&& d=4\ (\mbox{AdS}_5):  \no
&& T_{uu}=\frac{M\alpha ^4}{\pi  G_N R \left(\alpha ^2+u^2\right)^3 \left(\alpha ^2+v^2\right)},
T_{uv}=\frac{M\alpha ^4}{2 \pi  G_N R \left(\alpha ^2+u^2\right)^2 \left(\alpha ^2+v^2\right)^2} , \no
 && T_{vv}=\frac{M\alpha ^4}{\pi  G_N R \left(\alpha ^2+u^2\right) \left(\alpha ^2+v^2\right)^3} , \no
&&  T_{\theta \theta}=\frac{M\alpha ^4 (u-v)^2}{4 \pi  G_N R \left(\alpha ^2+u^2\right)^2 \left(\alpha ^2+v^2\right)^2} ,\ \ T_{\phi \phi}=\frac{M\alpha ^4 \sin ^2 \theta (u-v)^2}{4 \pi  G_N R \left(\alpha ^2+u^2\right)^2 \left(\alpha ^2+v^2\right)^2} .
\ea
The angular coordinates $\theta$ for $d=3$ and $(\theta,\phi)$ for $d=4$ are those of the
polar coordinates of $R^2:\ d\rho^2+\rho^2d\theta^2$ and $R^3:\ d\rho^2+\rho^2(d\theta^2+\sin^2\theta d\phi^2)$, respectively. For $d=4$, the above expression agrees with the one in \cite{HoIt}.

Note that we can easily confirm the traceless condition
\be
T_{ab}\eta^{ab}=0,
\ee
and the conservation law
\be
\de_a T^{ab}=0.
\ee

In particular, the energy density $T_{tt}$ in each dimension is
given by \ba && d=2 : \ \ T_{tt}=\f{M\ap^2}{4\pi G_N
R}\cdot\f{(t^2+x^2+\ap^2)^2+4t^2x^2}
{\left((x^2-t^2-\ap^2)^2+4x^2\ap^2\right)^2}, \no && d=3: \ \
T_{tt}=\f{M \ap^3}{\pi G_N R}\cdot
\f{(\rho^2+t^2+\ap^2)^2+2\rho^2t^2}{((\rho^2-t^2-\ap^2)^2+4\ap^2\rho^2)^{\f{5}{2}}},\no
&& d=4: \ \ T_{tt}=\frac{M\alpha^4}{\pi  G_N R }\cdot
\f{3(\rho^2+t^2+\ap^2)^2+4\rho^2t^2}
{((\rho^2-t^2-\ap^2)^2+4\ap^2\rho^2)^{3}} . \label{Ttt} \ea By using
these expressions we can confirm \be \int d^{d-1}x
T_{tt}=\f{mR}{\ap}=E, \ee for any $d$, by using the relation
(\ref{mrel}). Therefore the total energy agrees with that of the
particle (\ref{energy}) as expected.

\begin{figure}[ttt]
   \begin{center}
      \includegraphics[height=3cm]{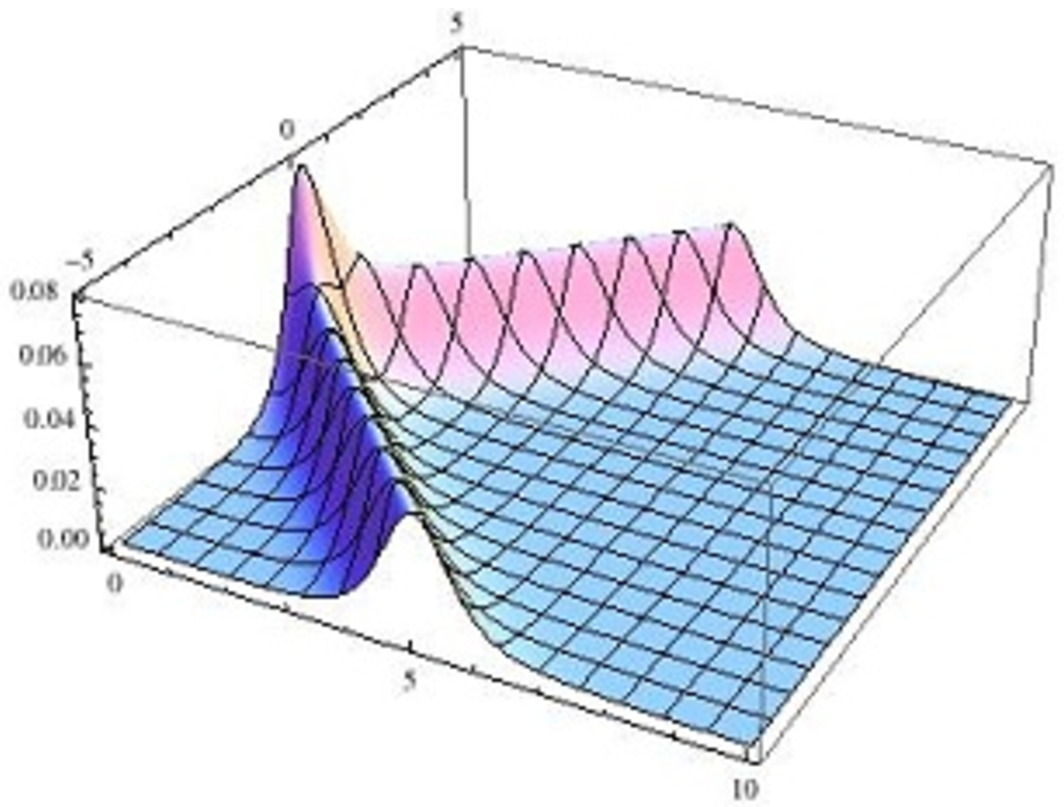}
      \includegraphics[height=3cm]{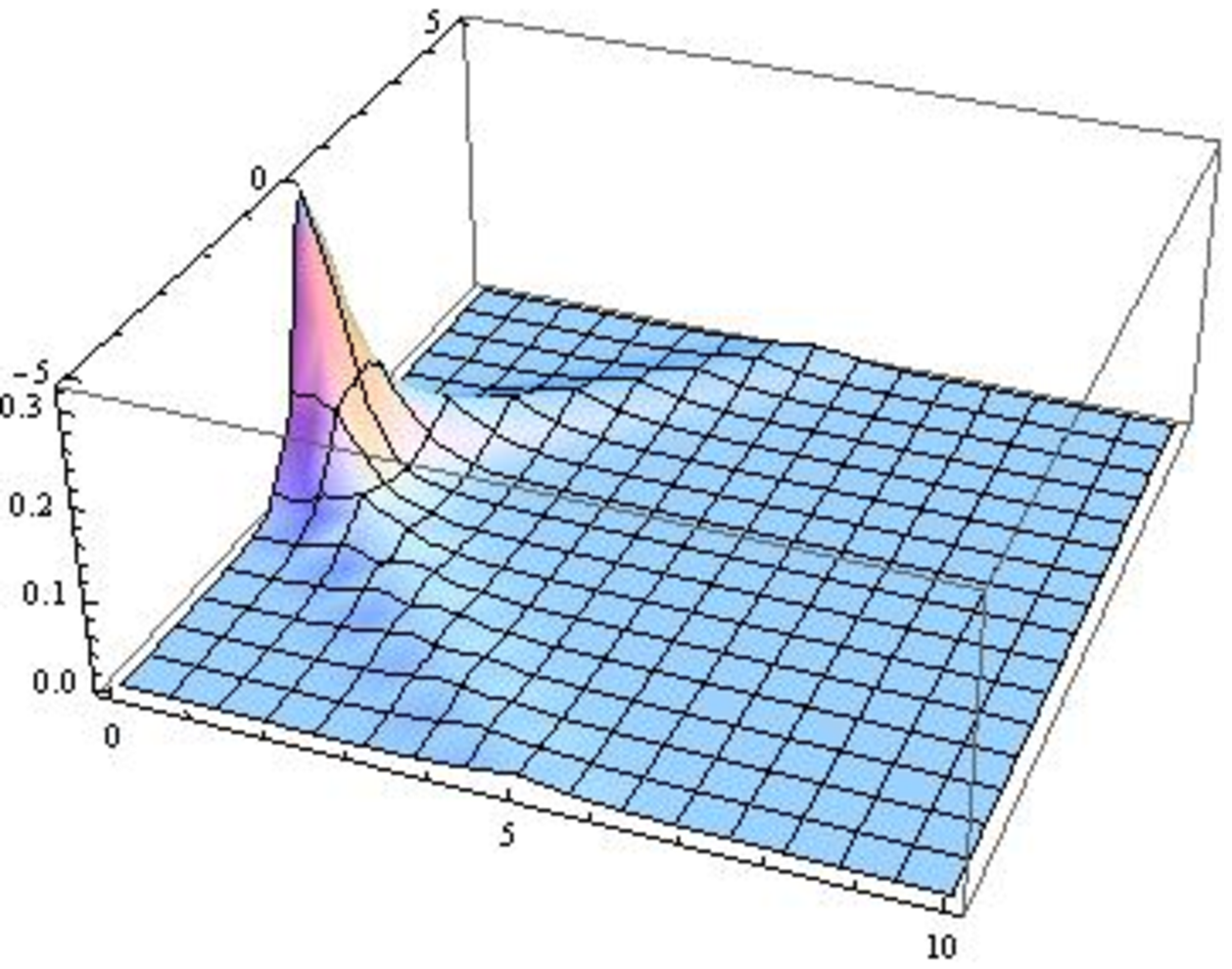}
      \includegraphics[height=3cm]{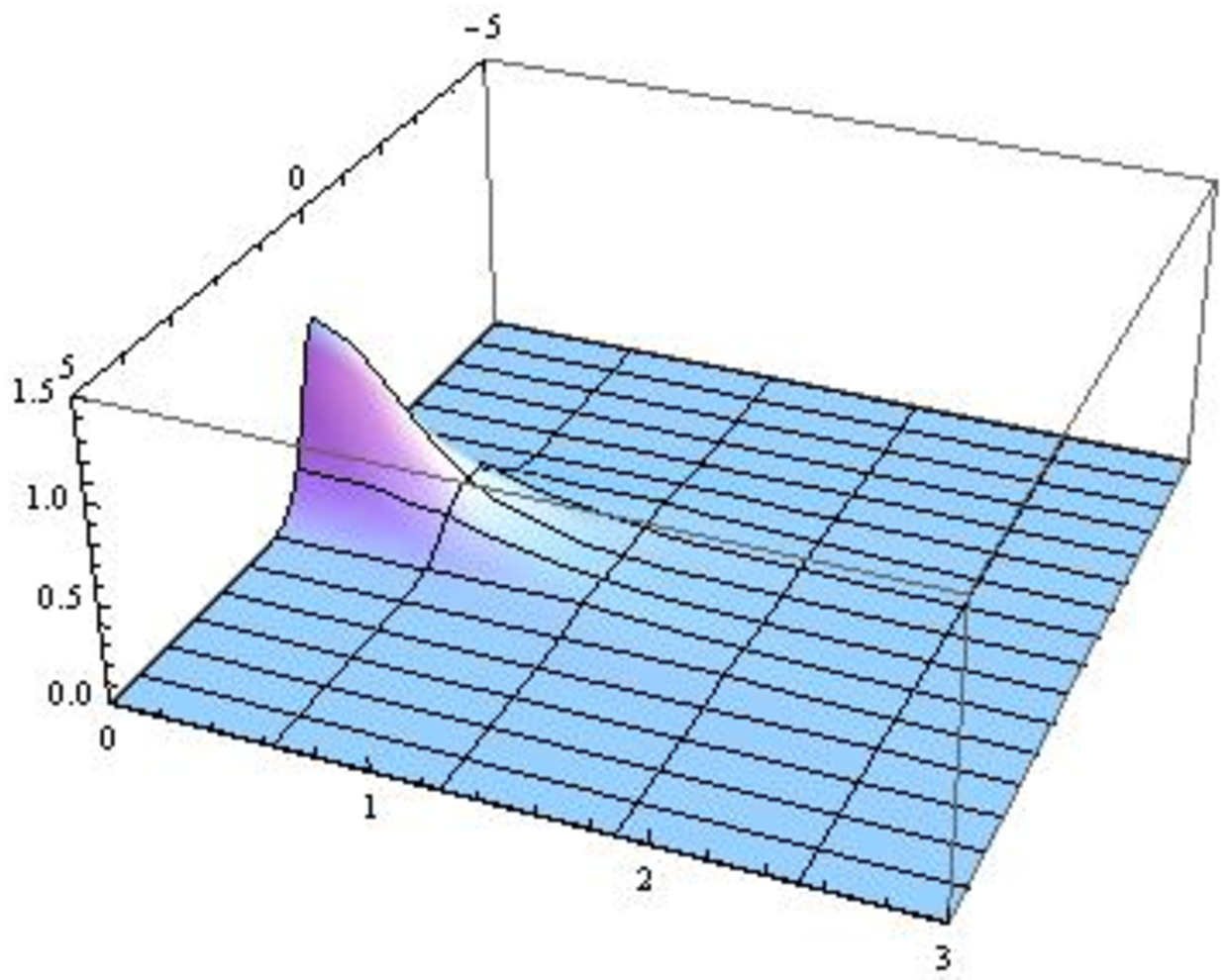}
   \end{center}
   \caption{The profiles of the energy density $T_{tt}$ for $d=2$ (left), $d=3$ (middle) and $d=4$ (right) as a function of $\rho$ and $t$ within the range
   $-5<t<5$.
   We set $\ap=R=M=G_N=1$.
    }\label{fig:EMtime}
\end{figure}

To understand how the excitations propagate, it is useful to look at
the time evolution of $T_{tt}$. This is plotted in
Fig.\ref{fig:EMtime}. We can observe a peak on the light-cone
$t^2=\rho^2$ as can be understood as the light-like propagation (or
shock waves as in \cite{HoIt}) of the initial excitations at
$t=\rho=0$.

\begin{figure}[ttt]
   \begin{center}
     \includegraphics[height=6cm]{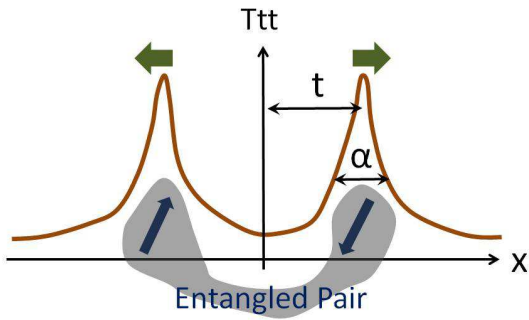}
   \end{center}
   \caption{A sketch of time evolution of energy density and entangled pairs for $d=2$.
   The understanding of detailed structure of quantum entanglement and entangled pairs is the main
   subject of this paper as will be studied in later sections.}\label{fig:pair}
\end{figure}

The parameter $\ap$ parameterizes the width of this peak and
therefore it measures the size of the localized excitations. Notice
that for $d=2$, the heights of the two peaks are equal and stay
constant due to the total energy conservation. We sketched the
essence of this behavior in Fig.\ref{fig:pair} for $d=2$. In the
zero width limit $\ap\to 0$ we find $T_{tt}$ gets delta-functionally
localized: \ba && d=2:\ \ T_{tt}\to
\f{E}{2}\left(\delta(t+\rho)+\delta(t-\rho)\right),\no && d=3:\ \
T_{tt}\to \f{E}{2\pi
\rho}\left(\delta(t+\rho)+\delta(t-\rho)\right),\no && d=4:\ \
T_{tt}\to \f{E}{4\pi
\rho^2}\left(\delta(t+\rho)+\delta(t-\rho)\right). \ea

\section{Perturbative Analysis of Holographic Entanglement Entropy under Local Quenches}

The entanglement entropy $S_A$ is defined as the von-Neumann entropy when we trace out
the subsystem $B$, which is the complement of $A$. The subsystem $A$ is an arbitrary chosen
space-like region on a time slice. Therefore, in a time-dependent background, $S_A$ depends on the time even if we fix the shape of the region $A$. This is expected to be an important quantity which characterizes various non-equilibrium processes in quantum many-body systems such as quantum quenches \cite{CaCaG,CaCaL,Eisler}.

In holographic setups, we can calculate $S_A$ by using the
holographic entanglement entropy (HEE). In $d+1$ dimensional static
gravity backgrounds, we can calculate $S_A$ by the formula \cite{RT}
\be S_A=\f{\mbox{Area}(\gamma_A)}{4G_N}, \label{HEE} \ee where
$\gamma_A$ is the $d-1$ dimensional minimal area surface on a time
slice which ends on the boundary of $A$ i.e. $\de A$. For
spherically symmetric subsystems, we can manifestly prove
(\ref{HEE}) in pure AdS spaces via the bulk to boundary relation
\cite{GKPW} as shown in \cite{CHM}. In time-dependent backgrounds,
there is no natural time-slice in the bulk AdS and we need to employ
the covariant version of HEE \cite{HRT}. This is given by redefining
$\gamma_A$ to be an extremal surface in the Lorentzian spacetime.

 Below we would like to compute the holographic entanglement entropy
 in the gravity dual of the local quench obtained from the coordinate
 transformation of the back hole solution (\ref{sbh}) as explained
 in the previous section. In this section we perform a perturbative
 calculation assuming that the back-reaction from the particle is very small,
keeping only the leading term proportional to $M$. This allows
us a relatively simple calculation which is applicable to any dimension $d$.

\subsection{Perturbative Calculations of HEE under Local Quenches}

Consider the first order perturbation of the metric
\be
g_{\mu\nu}=g^{(0)}_{\mu\nu}+g^{(1)}_{\mu\nu}+O(M^2),
\ee
where $g^{(0)}_{\mu\nu}$ represents the metric of pure AdS$_{d+1}$ (\ref{ads});
$g^{(1)}_{\mu\nu}$ is the leading perturbation due to the back-reaction, which is of
order $M$ in our case. We obtained $g^{(1)}_{\mu\nu}$ from the direct calculation
explained in the previous section, though we will not write its complicated
expression explicitly.

If we know the extremal surface $\gamma_A$ in the pure AdS$_{d+1}$,
then the perturbed area of an extremal surface can be found as \be
\Delta \mbox{Area}=\f{1}{2}\int
d^{d-1}\xi\s{G^{(0)}}\mbox{Tr}[G^{(1)}(G^{(0)})^{-1}]. \label{dea}
\ee Here $G$ is the induced metric on the surface $\gamma_A$: \be
G^{(0)}_{\ap\beta}=\f{\de X^\mu}{\de \xi^\ap}\f{\de X^\nu}{\de
\xi^\beta}g^{(0)}_{\mu\nu},\ \ \ \ \ G^{(1)}_{\ap\beta}=\f{\de
X^\mu}{\de \xi^\ap}\f{\de X^\nu}{\de \xi^\beta}g^{(1)}_{\mu\nu}, \ee
where $\xi^\ap\ \ (\ap=1,2,\ddd,d-1)$ is the coordinate of the
codimension two surface $\gamma_A$. The embedding function
$X^\mu(\xi)$ is that of $\gamma_A$ in the pure AdS$_{d+1}$. Notice
the useful fact that we do not need to know the precise shape of
 $\gamma_A$ in the perturbed metric to calculate (\ref{dea}).

\subsection{Explicit Calculations of HEE under Local Quenches}

Consider the $d=3$ case (AdS$_4$) first. We take the subsystem $A$
to be a round disk
 with the radius $l$, defined by $x_1^2+x_2^2\leq l^2$. The corresponding extremal surface
 (or equally minimal surface) is given by
 the half sphere \cite{RT}
\be
z=\s{l^2-x_1^2-x_2^2}. \label{mins}
 \ee
We can take $(\xi^1,\xi^2)=(x_1,x_2)$ and then we find that the area density is given by
\be
\f{1}{2}\s{\ti{G}^{(0)}}\mbox{Tr}[\ti{G}^{(1)}(\ti{G}^{(0)})^{-1}]
=\f{4MR^2\rho^2}{l\left(e^{2\beta}R^4+e^{-2\beta}(l^2-t^2)^2+2R^2(2\rho^2+t^2-l^2)\right)^{3/2}},
\label{dra}
\ee
where $\rho=\s{\sum_{i=1}^{d-1}x_i^2}$.

The area perturbation is found by integrating (\ref{dra}) over $x$
and $y$. By plugging this into (\ref{HEE}) we obtain the increased
amount of HEE, denoted by $\Delta S_A$, compared with $S_A$ for the
ground state dual to the pure AdS: \be \Delta S_A=\f{\pi M}{4G_N
R\ap
l}\left(\f{l^4-2l^2t^2+(\ap^2+t^2)^2}{\s{l^4+2l^2(\ap^2-t^2)+(\ap^2+t^2)^2}}
-|t^2+\ap^2-l^2|\right), \label{parea4} \ee where we used the
relation (\ref{aprel}). Note that the area law divergence
\cite{Area} is canceled in $\Delta S_A$ because $\Delta S_A$ is
defined from $S_A$ by subtracting the entanglement entropy for the
ground state. For example, at $t=0$ we find \be \Delta
S_A|_{t=0}=\f{\pi M l^3}{2G_N R\ap(l^2+\ap^2)}\ \ \ (l\leq \ap),\ \
\ \ \ \ \Delta S_A|_{t=0}=\f{\pi M \ap^3}{2G_N Rl(l^2+\ap^2)}\ \ \
(l>\ap), \ee

\begin{figure}[ttt]
   \begin{center}
     \includegraphics[height=3cm]{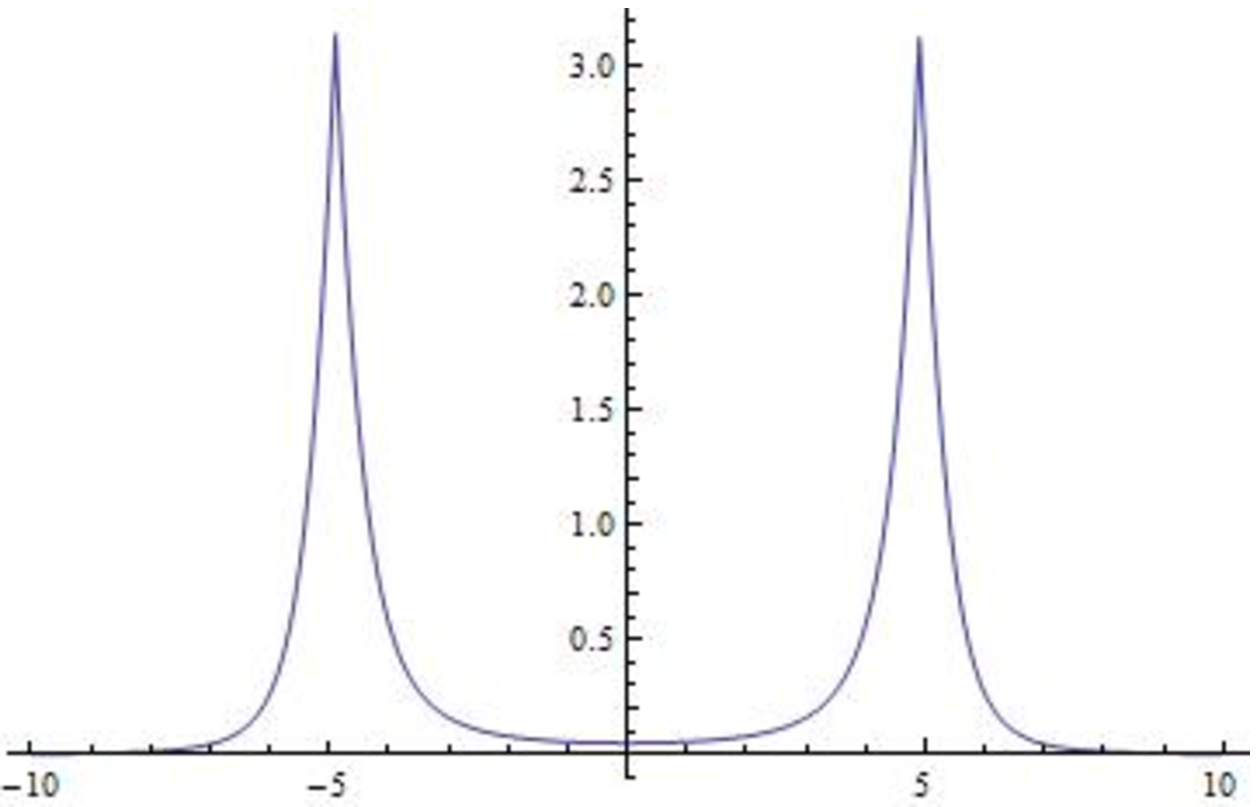}
     \includegraphics[height=3cm]{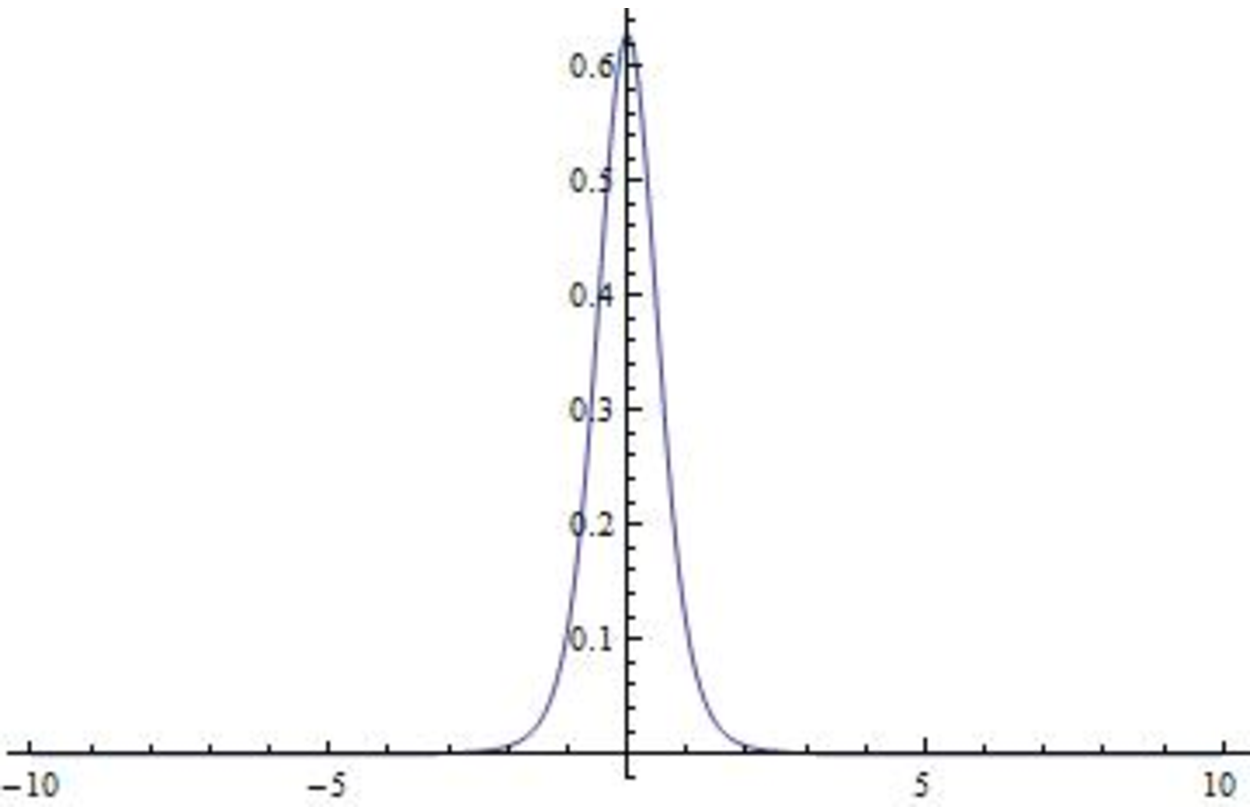}
     \includegraphics[height=3cm]{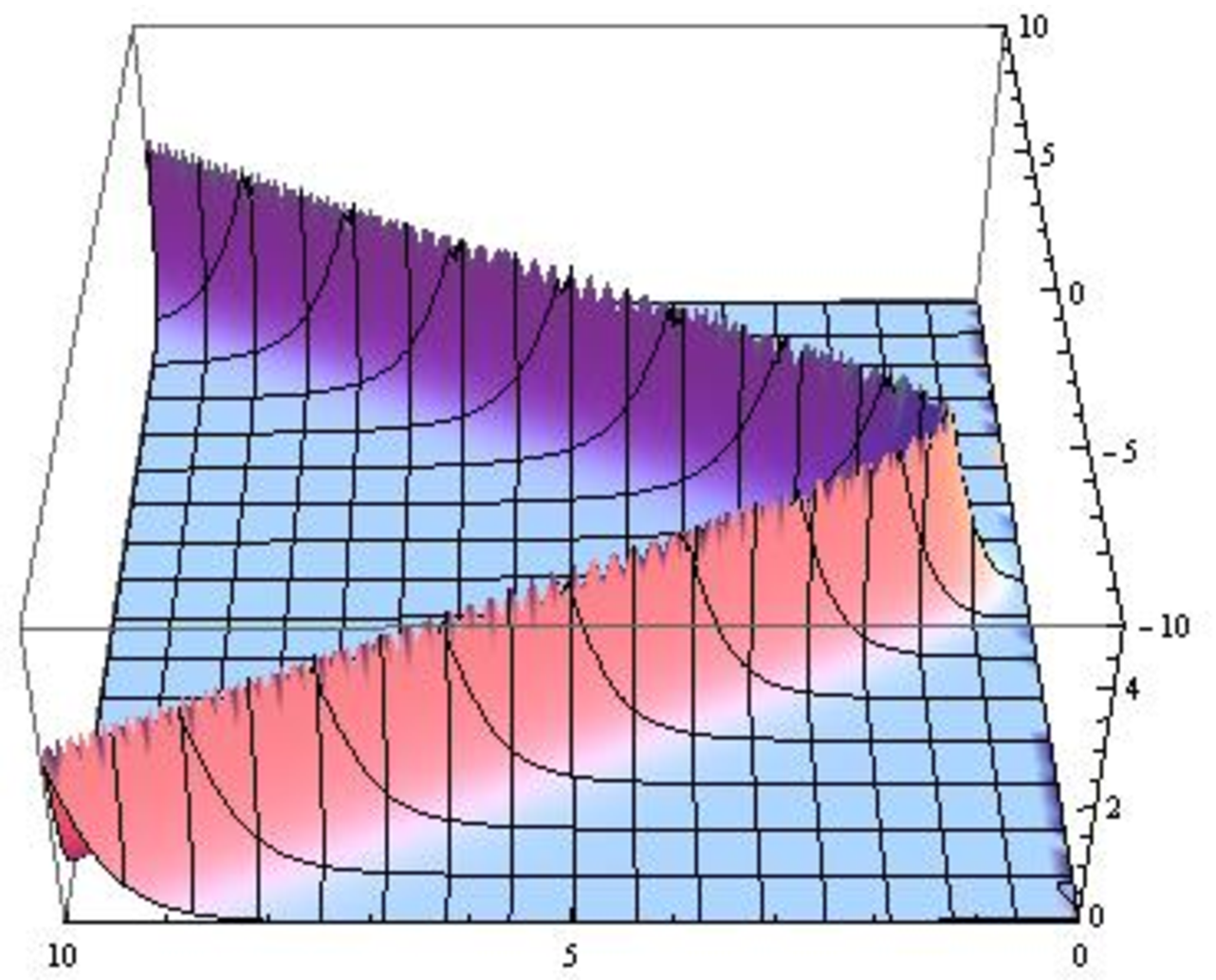}
   \end{center}
   \caption{The plots of $\Delta S_A$ at $\xi=0$ for $d=3$. The left and middle one describe
   $S_A$ as a function of $t$. We choose $(\ap,l)=(1,5)$ (left) and $(\ap,l)=(1,0.5)$ (middle), respectively. The right 3d graph expresses $S_A$ as a function of $t$ and $l$ for $\ap=1$.
   The horizontal
   coordinate corresponds to $l$. We took the range $-10<t<10$ and $0<l<10$. We set $R=4G_N=M=1$.
    }\label{fig:timed}
\end{figure}

The time-evolution of (\ref{parea4}) is plotted in
Fig.\ref{fig:timed}. Notice that $\Delta S_A$ respects the
time-reversal symmetry. When $l>\ap$, the entanglement entropy
initially grows with the time for $t>0$ and reaches the maximum at
$t=\s{l^2-\ap^2}$ as in the left graph in Fig.\ref{fig:timed}.
Later, it decreases as \be \Delta S_A\simeq \f{\pi M \ap^3l^3}{2G_N
Rt^6}.\label{decayt} \ee When $l<\ap$, the entanglement entropy
always decreases for $t>0$. At late time we have the behavior
(\ref{decayt}). Notice also that the width $\Delta t$ of the peak
around $t\simeq \s{l^2-\ap^2}$ is estimated as $\Delta t\sim \ap$
when $\ap<<l$. These results can be intuitively understood because
at $t=\s{l^2-\ap^2}$ the particle in AdS passes through the  minimal
surface (\ref{mins}). Refer to the Fig.\ref{fig:setup} again. In the
dual CFT, they can be naturally understood if we remember the
excitations propagate at the speed of light as we will discuss
later.

Notice also an interesting property that the height of the peak at $t=\s{l^2-\ap^2}$
stays constant, given by $S_A=\f{\pi M}{4G_N R}$,
and thus does not decrease under the time-evolution. This can also be seen from
the right graph in Fig.\ref{fig:timed}. This is rather different from the behavior of
energy stress tensor studied in the previous section (see Fig.\ref{fig:EMtime}).

It is also intriguing to shift the center of the subsystem $A$ relative to the trajectory of the falling particle in the $(x_1,y_2)$ plane. We take the minimal surface in the pure AdS
to be
\be
z=\s{l^2-(x_1-\xi)^2-x_2^2},
\ee
where $\xi$ is the distance between the center of $A$ and the falling particle.

\begin{figure}[ttt]
   \begin{center}
     \includegraphics[height=3cm]{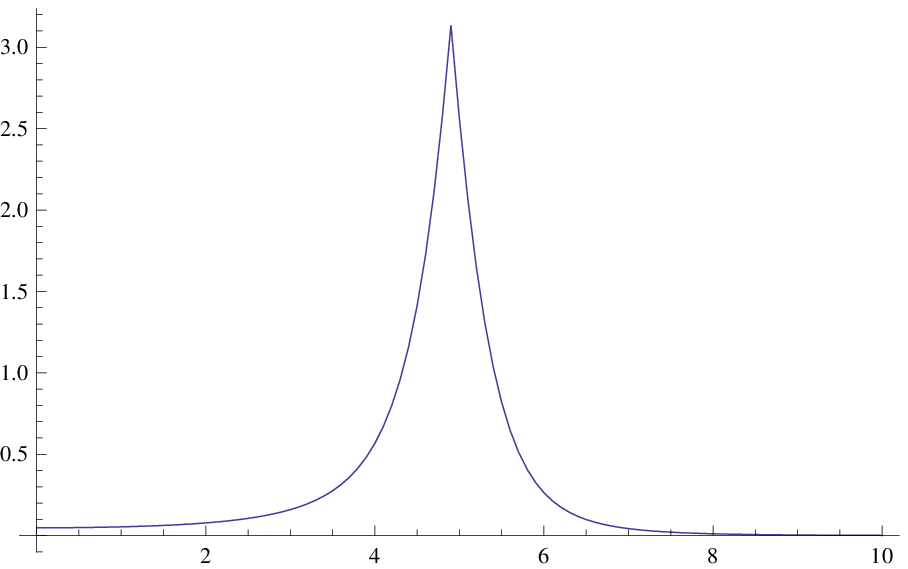}
     \includegraphics[height=3cm]{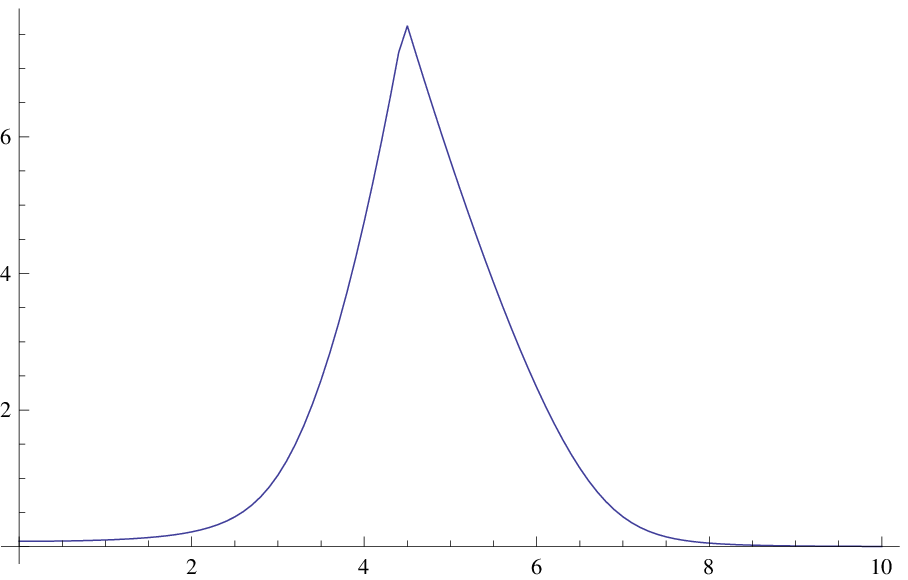}
     \includegraphics[height=3cm]{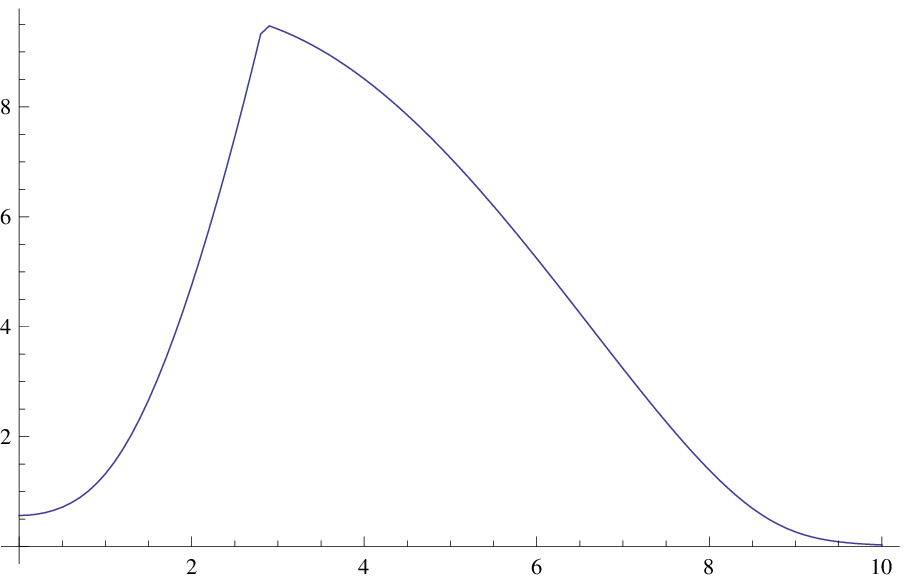}
   \end{center}
   \caption{The plots of $\Delta S_A$ for $d=3$ as a function of time $t$. The left, middle and right graph correspond to $(\ap,l,\xi)=(1,5,0),(1,5,2)$ and $(1,5,4)$, respectively. We set $R=4G_N=M=1$.
    }\label{fig:timeds}
\end{figure}

We plotted the results of $\Delta S_A$ in Fig.\ref{fig:timeds} for different values of
$\xi$. In general one may notice that the peak is broadened so that it is spread over
$l-\xi <t<l+\xi$. In the gravity dual, this is easily explained from the propagation of
gravitational waves from the falling particle. This also qualitatively agrees with the results of local quenches in two dimensional CFTs \cite{CaCaL}.

It is straightforward to generalize the above results to other dimensions.
In $d=2$ and $d=4$, the expression of $\Delta S_A$ (\ref{parea4}) for $d=3$ is replaced with\footnote{In $d=2$ the subsystem is chosen to be an interval $-l\leq x\leq l$.}
\ba
&& d=2:\ \ \  \Delta S_A=\f{2Ml\ap+M(l^2-\ap^2-t^2)\arctan\left(\f{2\ap l}{t^2+\ap^2-l^2}\right)}{8G_N lR \ap},\label{peet}\\
&& d=4:\ \ \ \Delta S_A =\frac{\pi  M}{8G_N l R\ap} \left(\frac{1}{\frac{2 \alpha  l}{\left(\alpha ^2-l^2+t^2\right)^2}+\frac{1}{2 \alpha  l}}-3 \left(\alpha ^2-l^2+t^2\right) \arctan
   \left(\frac{2 \alpha  l}{\alpha ^2-l^2+t^2}\right)+4 \alpha  l\right). \nonumber
\ea
We can confirm that the behaviors of $\Delta S_A$ in $d=2$ and $d=4$ are very similar to
that in $d=3$ and thus we will not show them the explicitly.

At late time $t>>l$, they are approximated by
\ba
&& d=2:\ \ \ \Delta S_A\simeq \f{M\ap^2l^2}{3G_N Rt^4},\no
&& d=4:\ \ \ \Delta S_A \simeq \frac{8 \pi  \alpha^4  l^4 M}{5G_NR t^8}.
\ea
This shows that $\Delta S_A$ decays like $\sim l^d\cdot t^{-2d}$ at late time. Notice that the perturbative calculation is
always justified at late time as the back-reaction to $\gamma_A$ for a finite $l$ clearly
gets suppressed.

\subsection{Small Subsystem Limit: An Analogue of the first law of thermodynamics}

In the small limit of the subsystem size $l$, we can trust the
perturbative results of $\Delta S_A$ (\ref{parea4}) and
(\ref{peet}). This is because the surface $\gamma_A$ is situated
near the AdS boundary and therefore the deviation of the metric from
the pure AdS is very small \cite{BNTU}. In asymptotically AdS
backgrounds which are static and translation invariant, a relation
which looks like the first law of thermodynamics has been found in
the small size limit of $A$ \cite{BNTU}: \be T_{eff}\cdot \Delta
S_A=\Delta E_A, \label{firstEE} \ee where $E_A$ is the energy in the
subsystem $A$ given by $E_A=\int_A dx^{d-1}T_{tt}$. The effective
temperature is defined by $T_{eff}=\f{c_{A}}{l}$, where the constant
$c_A$ only depends on the shape of the subsystem $A$ and is
independent from the details of the CFT we consider. When $A$ is a
$d-1$ dimensional ball with the radius $l$ as we choose in this
paper, we have \be T_{eff}=\f{d+1}{2\pi l}. \ee

Thus it is intriguing to see if this relation (\ref{firstEE}) holds in our time-dependent and inhomogeneous
setup of local quenches. We can calculate the energy density from (\ref{Ttt}), and $\Delta
S_A$ from (\ref{parea4}) and (\ref{peet}) in the limit $l<<\s{\ap^2+t^2}$ as follows:
\ba
&& d=2: \ \ T_{tt}=\f{M\ap^2}{4\pi G_N R(t^2+\ap^2)^2}, \ \ \
\Delta S_A=\f{M\ap^2 l^2}{3G_N R(t^2+\ap^2)^2}, \no
&&  d=3: \ \ T_{tt}=\f{M\ap^3}{\pi G_N R(t^2+\ap^2)^3}, \ \ \
\Delta S_A=\f{\pi \ap^3 l^3}{2G_N R (t^2+\ap^2)^3}, \no
&&  d=4: \ \ T_{tt}=\f{3M\ap^4}{\pi G_N R(t^2+\ap^2)^4}, \ \ \
\Delta S_A=\f{8\pi M\ap^4 l^4}{5G_N R (t^2+\ap^2)^4}.
\ea
By using these expressions we can explicitly confirm the relation (\ref{firstEE}) for any $d$.

\subsection{Large Subsystem Limit}

Before we go on, we would like to write down the result in the large subsystem limit $l>>\s{t^2+\ap^2}$. In this limit, we find from (\ref{parea4}) and (\ref{peet}):
\ba
&& d=2: \ \ \Delta S_A=\f{8\ap^2}{3l^2}mR,\no
&& d=3: \ \ \Delta S_A=\f{\pi\ap^3}{l^3}mR,\no
&& d=4: \ \ \Delta S_A=\f{64\ap^4}{5l^4}mR.  \label{sapert}
\ea
Note that these are time-independent.

\section{Exact Holographic Entanglement Entropy for $2$d Local Quenches}

We can actually find exact extremal surfaces (i.e. geodesics) in the
AdS$_3$ case ($d=2$). We take the subsystem $A$ to be an interval at
a constant time in the dual CFT$_2$. In the holographic calculation,
first we obtain a geodesic in the metric (\ref{sbh}), which is
asymptotically global AdS$_{3}$. Then we map it into an
asymptotically Poincare AdS metric by the coordinate transformation
(\ref{corf}).

Initially, we will assume $M-R^2<0$ and thus the geometry (\ref{sbh}) has the deficit angle at $r=0$. If we were allowed to take $\theta$ to be $2\pi \f{R}{\s{R^2-M}}$ instead of $2\pi$, the geometry gets smooth. Later we will come back to the case $M-R^2>0$ i.e.
the BTZ black hole.

In the near AdS boundary limit, the map between the point
$(\tau_{\infty},\theta_{\infty},r_{\infty})$ in (\ref{sbh}) and the
point $(t,z_{\infty},x_{\infty})$ in the Poincare AdS is given
by\footnote{To fix the signs of $\tau_\infty$ and $\theta_\infty$,
we need to go back to the original coordinate transformation
(\ref{corf}).} \ba &&
\tan\tau_{\infty}=\f{2Rt}{R^2e^{\beta}+e^{-\beta}(x_{\infty}^2-t^2))},\no
&&
\tan\theta_{\infty}=-\f{2Rx_{\infty}}{e^{-\beta}(x_{\infty}^2-t^2)-R^2e^{\beta}},\no
&& r_{\infty}=\f{1}{z_{\infty}}\s{R^2x_{\infty}^2+\f{1}{4}
\left(e^{-\beta}(x_{\infty}^2-t^2)-R^2e^{\beta}\right)^2 }. \ea
Notice that $z_\infty$ is interpreted as the UV cut off (or lattice
spacing) in the dual CFT. In the above expressions, we chose the
range of $\tau_\infty$ and $\theta_\infty$ to be $[-\pi,\pi]$.

We can specify the geodesic $\gamma_A$ in (\ref{sbh}) by \be
\tau=\tau(\theta),\ \ \ r=r(\theta). \ee Its length $|\gamma_A|$
reads \be |\gamma_A|=\int d\theta
\s{r^2+\f{R^2}{r^2+R^2-M}r'^2-(r^2+R^2-M)\tau'^2}. \label{areaf} \ee
The minimal length condition is summarized as \ba &&
\f{d\tau}{d\theta}=\f{A r^2}{r^2+R^2-M},\no &&
\f{dr}{d\theta}=\f{r}{R}\s{A^2 r^2+(B^2 r^2-1)(r^2+R^2-M)},
\label{difge} \ea where $A$ and $B$ are integration constants.

\subsection{Symmetric Intervals}

Consider the case where the subsystem $A$ is given by an interval
$-l\leq x\leq l$ at time $t$. The excitations of the local quench
occur at $x=0$ because the falling particle is situated at $x=0$. In
the global AdS, this is mapped into an interval $-\theta_{\infty}
\leq \theta\leq \theta_{\infty}$ at a constant time slice
$\tau=\tau_{\infty}$, where we can assume $0<\theta_{\infty}<\pi$
without losing generality (see Fig.\ref{fig:geodesic}). Since we can
assume $\tau$ is constant on $\gamma_A$, the extremal surface
condition (\ref{difge}) gets simplified to \be
\f{dr}{d\theta}=\f{r}{Rr_*}\s{(r^2+R^2-M)(r^2-r_*^2)}, \ee where
$r=r_*$ is an integration constant and $r_*$ is the minimum value of
$r$ on $\gamma_A$, i.e. the turning point.

\subsubsection{Case 1:  $M\leq R^2$}

\begin{figure}[ttt]
   \begin{center}
     \includegraphics[height=6cm]{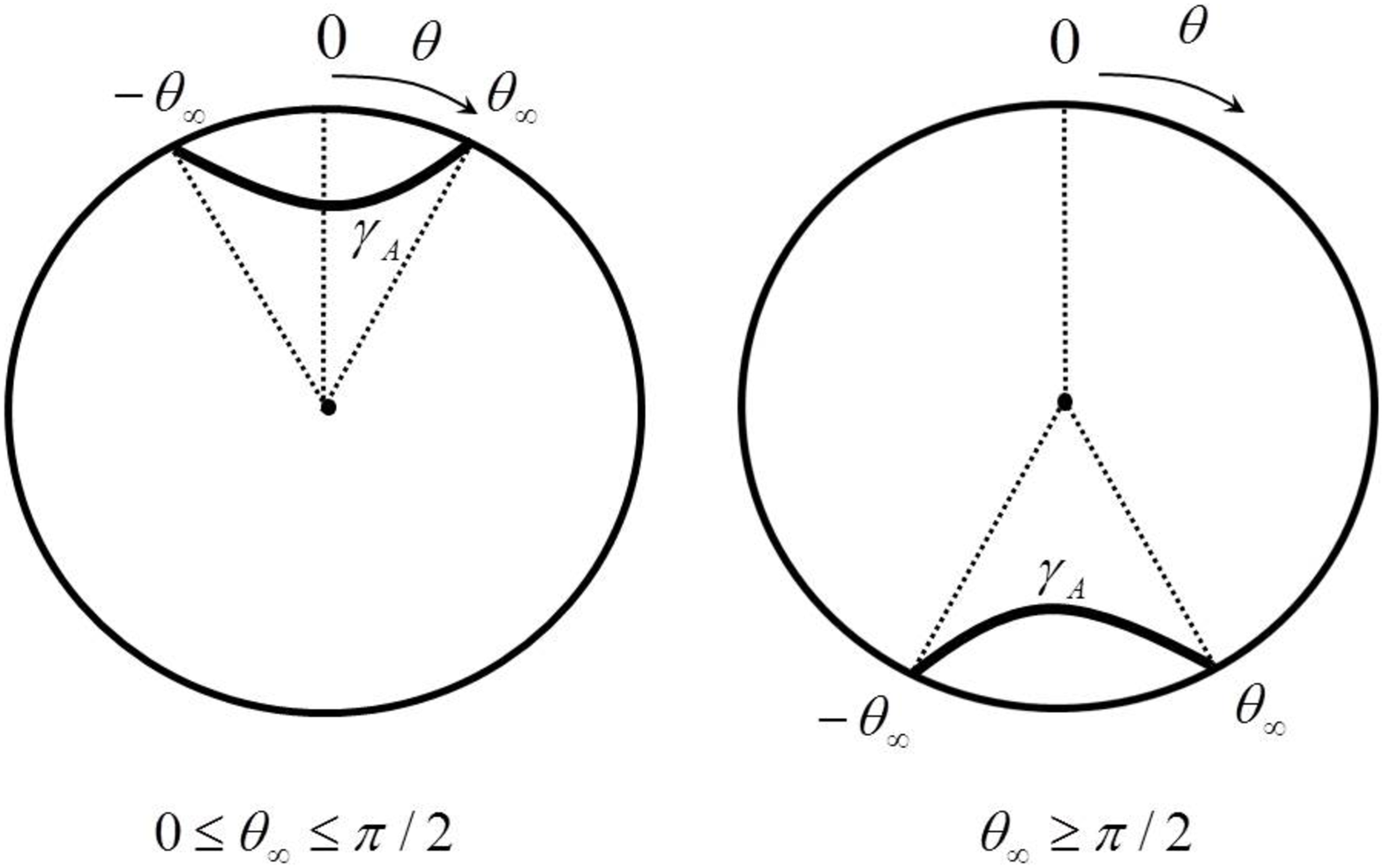}
   \end{center}
   \caption{The sketch of $\gamma_A$ in the asymptotically global AdS background (\ref{sbh}).
   The left and right picture corresponds to $0\leq \theta_\infty\leq \pi/2$ and
   $\theta_\infty\geq \pi/2$, respectively.
    }\label{fig:geodesic}
\end{figure}

Let us first assume $0<\theta_{\infty}<\f{\pi}{2}$. The curve $\gamma_A$ with the minimum length is given by the geodesic which takes the angle range $-\theta_\infty\leq \theta<
 \theta_\infty$ (see the left picture of Fig.\ref{fig:geodesic}).
 The point $\theta=0$ corresponds to the turning point $r=r_*$.
Thus we find\footnote{In this paper, the function $\arcsin(x)$ takes values between
$-\f{\pi}{2}$ and $\f{\pi}{2}$.}
\ba
&& \theta_{\infty}=\int^{\infty}_{r_*}dr\f{Rr_*}{r\s{(r^2+R^2-M)(r^2-r_*^2)}} \no
&& \ \ \ \ = \f{R}{2\s{R^2-M}}\left[\f{\pi}{2}
+\arcsin\left(\f{R^2-M-r_*^2}{R^2-M+r_*^2}\right)\right].
\ea
It is useful to rewrite this into
\be
\cos\left(\f{2\s{R^2-M}}{R}\theta_{\infty}\right)=\f{r_*^2-R^2+M}{r_*^2+R^2-M}. \label{cosr}
\ee
Finally, the HEE is given by
\ba
&& S_A=\f{R}{2G_N}\int^{r_{\infty}}_{r_*}dr\f{r}{\s{(r^2+R^2-M)(r^2-r_*^2)}}\no
&& \ \ \ =\f{R}{2G_N}\log \f{2r_{\infty}}{\s{R^2-M+r_*^2}}. \label{finco}
\ea

In the case $\theta_{\infty}>\f{\pi}{2}$, on the other hand,
the curve $\gamma_A$ is given by the geodesic which takes the angular
 range $\theta_\infty<|\theta|<\pi$ (see the right picture of
 Fig.\ref{fig:geodesic}). Thus its turning
point is at $\theta=\pi$. This guarantees the basic property of
entanglement entropy written as $S_A=S_B$ for pure states, where $B$
is the complement of $A$. Thus for $\theta_{\infty}>\f{\pi}{2}$, we
find the correct HEE from (\ref{finco}) by replacing $\theta_\infty$
with $\pi-\theta_\infty$ in (\ref{cosr}).

\subsubsection{Case 2:  $M>R^2$}

For $M>R^2$, the geometry (\ref{sbh}) becomes the BTZ black hole.
In this case, the basic result can be obtained from the previous one by the analytic continuation $\s{R^2-M}\to i\s{M-R^2}$: the relation (\ref{cosr}) is replaced with
\be
\cosh \left(\f{2\s{M-R^2}}{R}\theta_\infty\right)=\f{r_*^2-R^2+M}{r_*^2+R^2-M}.
\ee
For $0<\theta_\infty<\f{\pi}{2}$, the holographic entanglement entropy $S_A$
can be found from the same expression
(\ref{finco}), which is denoted by $S_A(\theta_\infty)$. For $\theta_\infty>\f{\pi}{2}$, $S_A$ is given by $S_A(\pi-\theta_\infty)$ as in the $M<R^2$ case.

In this calculation we implicitly assume that we are considering a star solution and that outside of the star is described by the BTZ black hole solution (\ref{sbh}). Thus this is dual to a pure state
in the CFT and indeed our construction satisfies $S_A=S_B$.

\subsubsection{Final Results}

\begin{figure}[ttt]
   \begin{center}
     \includegraphics[height=3cm]{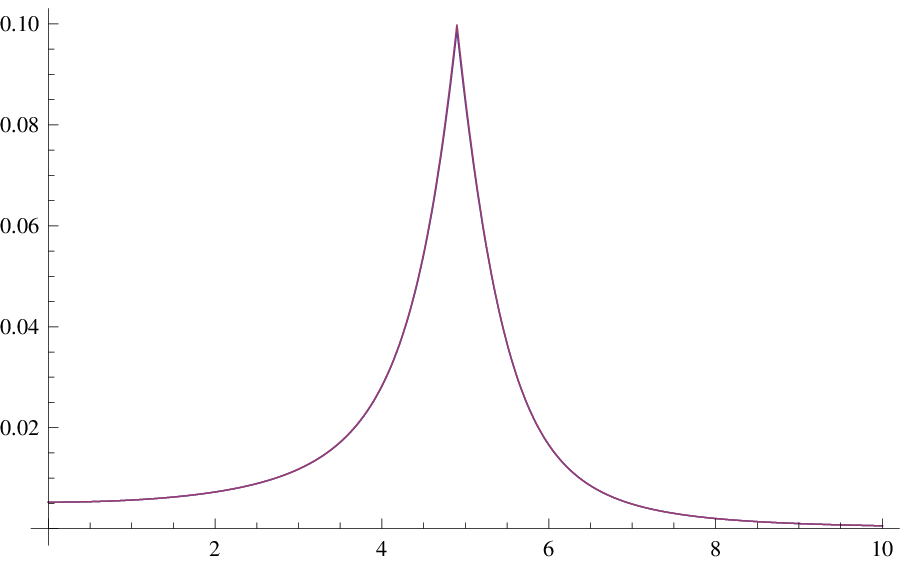}
     \includegraphics[height=3cm]{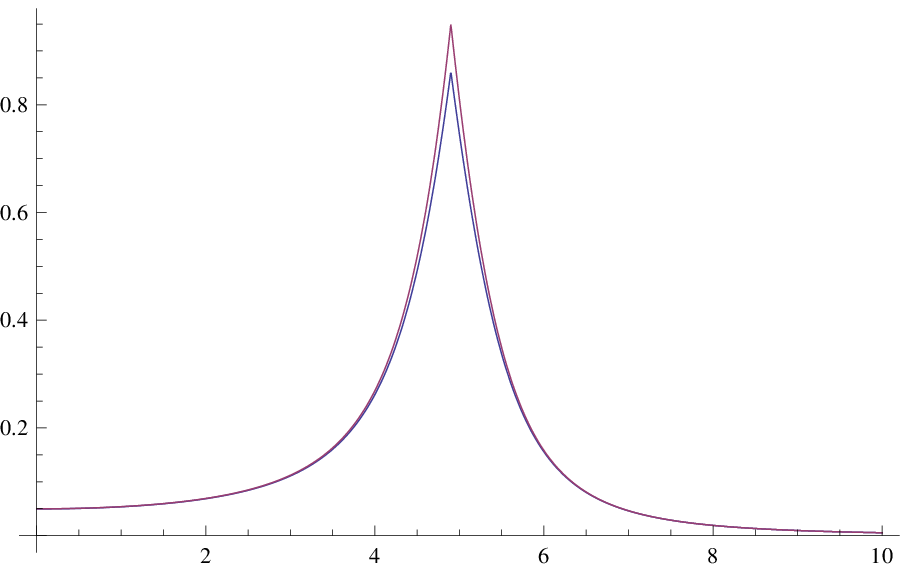}
     \includegraphics[height=3cm]{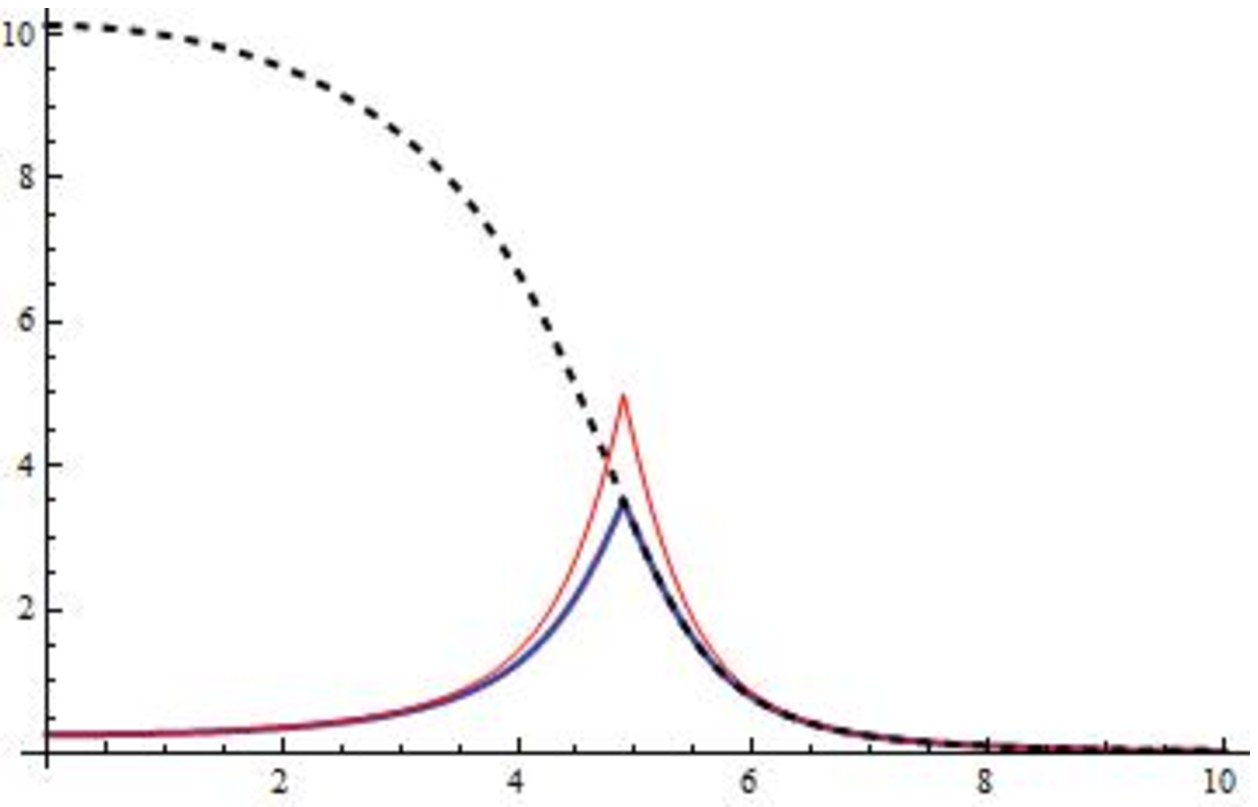}
   \end{center}
   \caption{The exact plots of $\Delta S_A$ as a function of time $t(>0)$. The left, middle and right graph correspond to $M=0.1,0.95,5$, respectively. The blue and red curve correspond to the result from the exact analysis and perturbation theory, respectively. They coincide almost completely for $M=0.1$. The dotted black one for $M=5$ corresponds to the thermal local quench given by (\ref{EEbh}). We set $R=\ap=4G_N=1$ and $l=5$.
    }\label{fig:timede}
\end{figure}

The final result is plotted in Fig.\ref{fig:timede}. For a small $M$ the exact result nicely agrees with the one from the perturbation theory (\ref{peet}). Even for a large $M$, the agreement is very good except the peak at $t\sim \s{l^2-\ap^2}$. Indeed, we can analytically show that in both the limit $t\to 0$ and
$t=\infty$, $S_A$ approaches to
\be
S_A(0)=S_A(t=\infty)=\f{c}{3}\log \f{2l}{z_{\infty}}, \label{frgr}
\ee
where we employed the well-known relation $c=\f{3R}{2G_N}$ \cite{BrHe}.
This reproduces the well-known result of the entanglement entropy for ground states in CFTs \cite{HLW,CaCa,CCreview}. Remember that $z_\infty$ corresponds to the UV cut off (lattice spacing) in the dual CFTs.

\subsubsection{Thermal Local Quenches}

In the previous analysis for $M>R^2$, we obtained $S_A$ by replacing
a black hole with a star in order to have a gravity dual of a pure
state, which is usually assumed in the study of quantum quenches. If
we deal with a falling black hole instead of a massive star, we will
obtain $S_A$ as follows \be S_A=\min \left\{S_A(\theta_\infty),\
S_A(\pi-\theta_\infty)+\f{\pi\s{M-R^2}}{2G_N}\right\}, \label{EEbh}
\ee where $\min\{a,b\}$ denotes the smaller one among $a$ and $b$.
The term $\f{\pi\s{M-R^2}}{2G_N}$ corresponds to a half of black
hole entropy and arises because $\gamma_A$ wraps a half of horizon.
The topological condition of $\gamma_A$ of HEE requires that the
subsystem $A$ should be homologous to $\gamma_A$. Therefore in the
presence of the black hole horizon, we cannot simply change the
geodesic from the one passing through $\theta=0$ to the one passing
through $\theta=\pi$. If we want to deform $\gamma_A$ in this way,
we need also to wrap $\gamma_A$ on the horizon as its disconnected
part. We plotted $S_A$ for this thermal local quench in the third
graph of Fig.\ref{fig:timede}.

\subsection{General Formulation}

Now let us extend the previous analysis to more general setups where
the subsystem $A$ is given by an arbitrary interval $l^{(1)}\leq
x\leq l^{(2)}$ at the time $t$. The surface $\gamma_A$ is defined by
the geodesic curve whose two end points are given by
$(x,t,z)=(l^{(i)},t,z_{\infty})\ \ (i=1,2)$ in the Poincare AdS3 and
equally by
$(\tau,\theta,r)=(\tau^{(i)}_{\infty},\theta^{(i)}_{\infty},r^{(i)}_{\infty})$
in the global AdS. Note that their relations are given by: \ba &&
\tan\tau^{(i)}_{\infty}=\f{2Rt}{R^2e^{\beta}+e^{-\beta}((l^{(i)})^2-t^2)},\no
&&
\tan\theta^{(i)}_{\infty}=-\f{2Rl^{(i)}}{e^{-\beta}((l^{(i)})^2-t^2)-R^2
e^{\beta}},\no &&
r^{(i)}_{\infty}=\f{1}{z_{\infty}}\s{R^2(l^{(i)})^2+\f{1}{4}
\left(e^{-\beta}((l^{(i)})^2-t^2)-R^2e^{\beta}\right)^2 }. \ea

Then the HEE reads \be S_A=\f{R}{4G_N}\sum_{i=1}^2
\left[\int^{r^{(i)}_{\infty}}_{r_*} dr\f{B
r}{\s{A^2r^2+(B^2r^2-1)(r^2+R^2-M)}}\right]. \ee

By integrating (\ref{difge}) we find
\ba
&& |\tau^{(2)}_\infty-\tau^{(1)}_\infty|=\f{R}{\s{R^2-M}}\left[\f{\pi}{2}
+\arcsin\left(\f{B^2(M-R^2)+A^2-1}{\s{\left(B^2(R^2-M)+A^2-1\right)^2+4B^2(R^2-M)}}\right)\right],\no
&& |\theta^{(2)}_\infty-\theta^{(1)}_\infty|=\f{R}{\s{R^2-M}}\left[\f{\pi}{2}
+\arcsin\left(\f{B^2(R^2-M)+A^2-1}{\s{\left(B^2(R^2-M)+A^2-1\right)^2+4B^2(R^2-M)}}\right)\right].
\nonumber \\
\ea
If we assume $0<|\theta^{(2)}_\infty-\theta^{(1)}_\infty|<\pi$, the HEE is computed as follows:
\ba
&& S_A=\f{R}{4G_N}\left[\log (r^{(1)}_{\infty}\cdot r^{(2)}_{\infty})+\log \left( \f{4B^2}{\s{\left(B^2(R^2-M)+A^2-1\right)^2+4B^2(R^2-M)}}\right)\right], \no
&& \ \ =\f{R}{4G_N}\left[\log (r^{(1)}_{\infty}\cdot r^{(2)}_{\infty})
+\log \f{2\cos\left[\f{\s{R^2-M}|\tau^{(2)}_\infty-\tau^{(1)}_\infty|}{R}\right]
-2\cos\left[\f{\s{R^2-M}|\theta^{(2)}_\infty-\theta^{(1)}_\infty|}{R}\right]}{R^2-M}\right]. \label{genints}
\ea
In the case $\Delta \theta_{\infty}=|\theta^{(2)}_\infty-\theta^{(1)}_\infty|>\pi$, we need to replace $\Delta \theta_{\infty} \to  2\pi-\Delta \theta_{\infty}$ as we did in the previous subsection.

So far we assumed $M<R^2$. If $M>R^2$, then the solution (\ref{sbh}) is a BTZ black hole without any deficit angle. However, our analytical calculation done before still holds via the analytical continuation as in the previous case.

\subsection{An Interval with An Excited End Point}

\begin{figure}[ttt]
   \begin{center}
     \includegraphics[height=3cm]{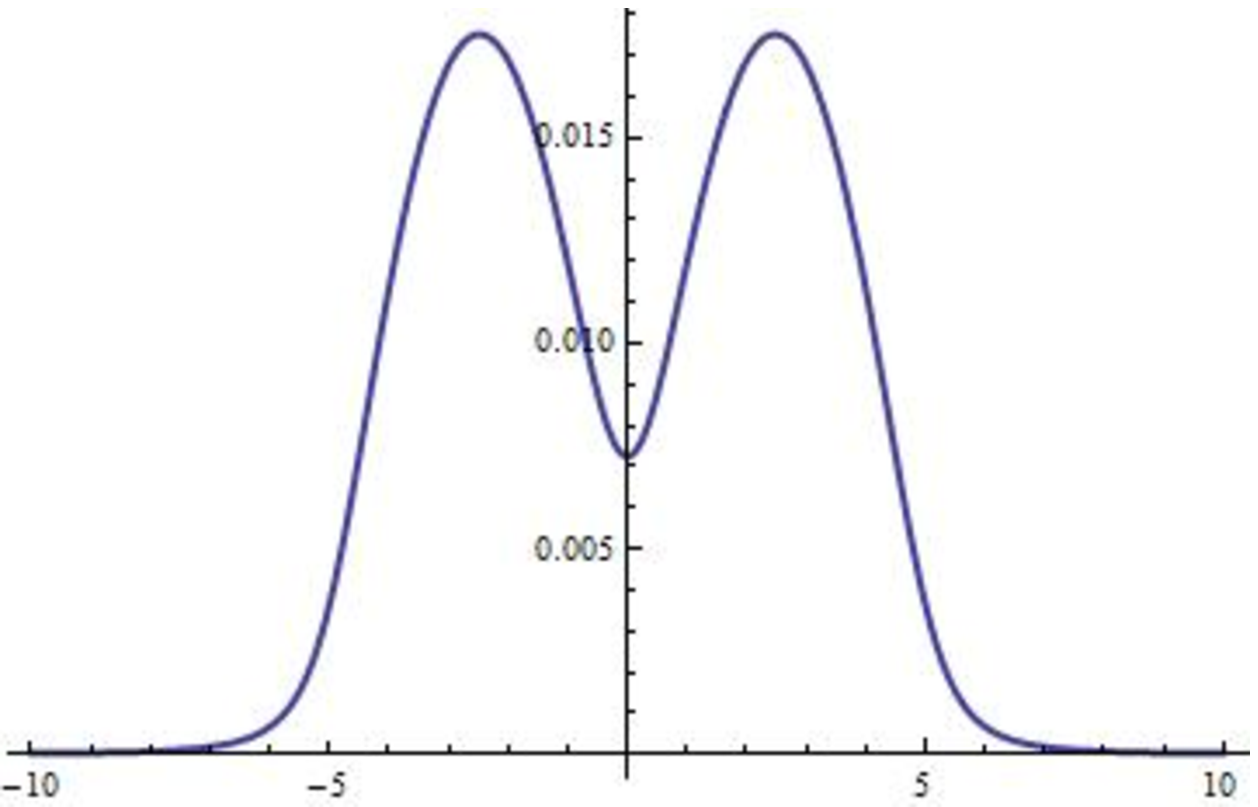}
     \includegraphics[height=3cm]{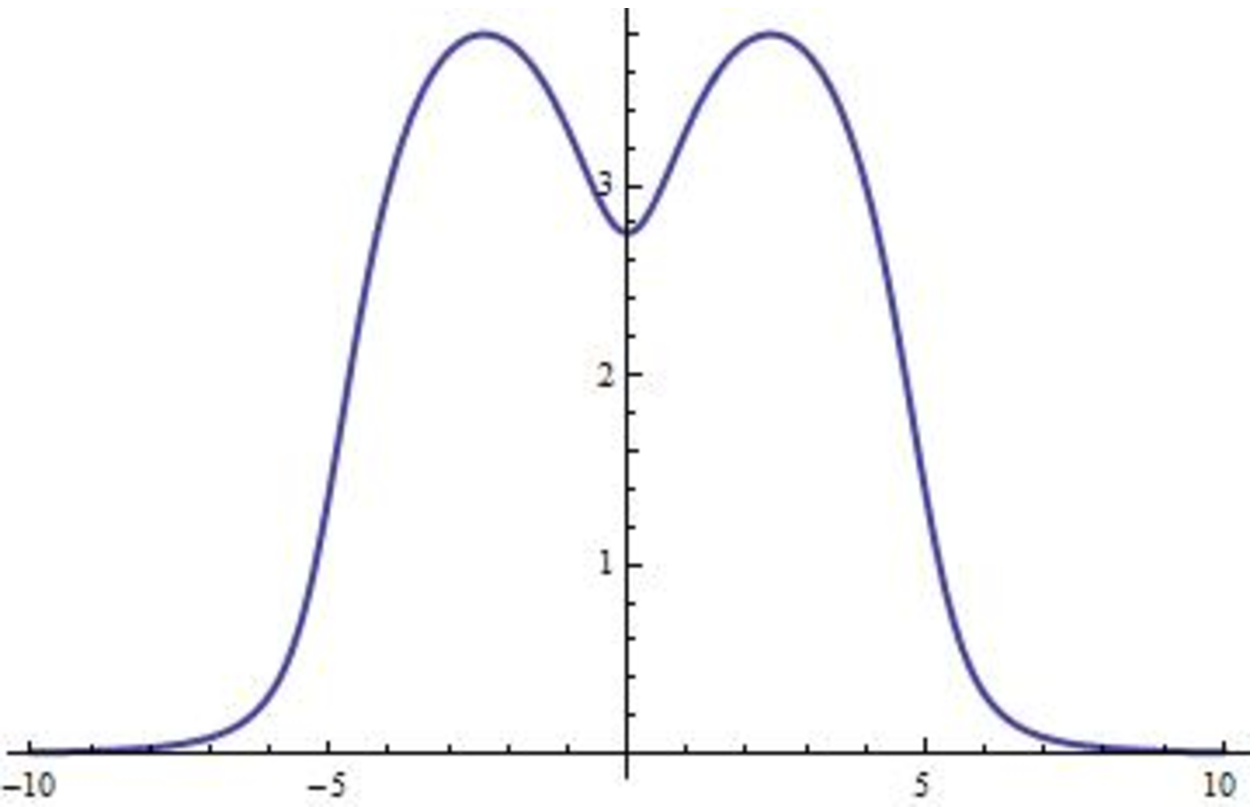}
     \includegraphics[height=3cm]{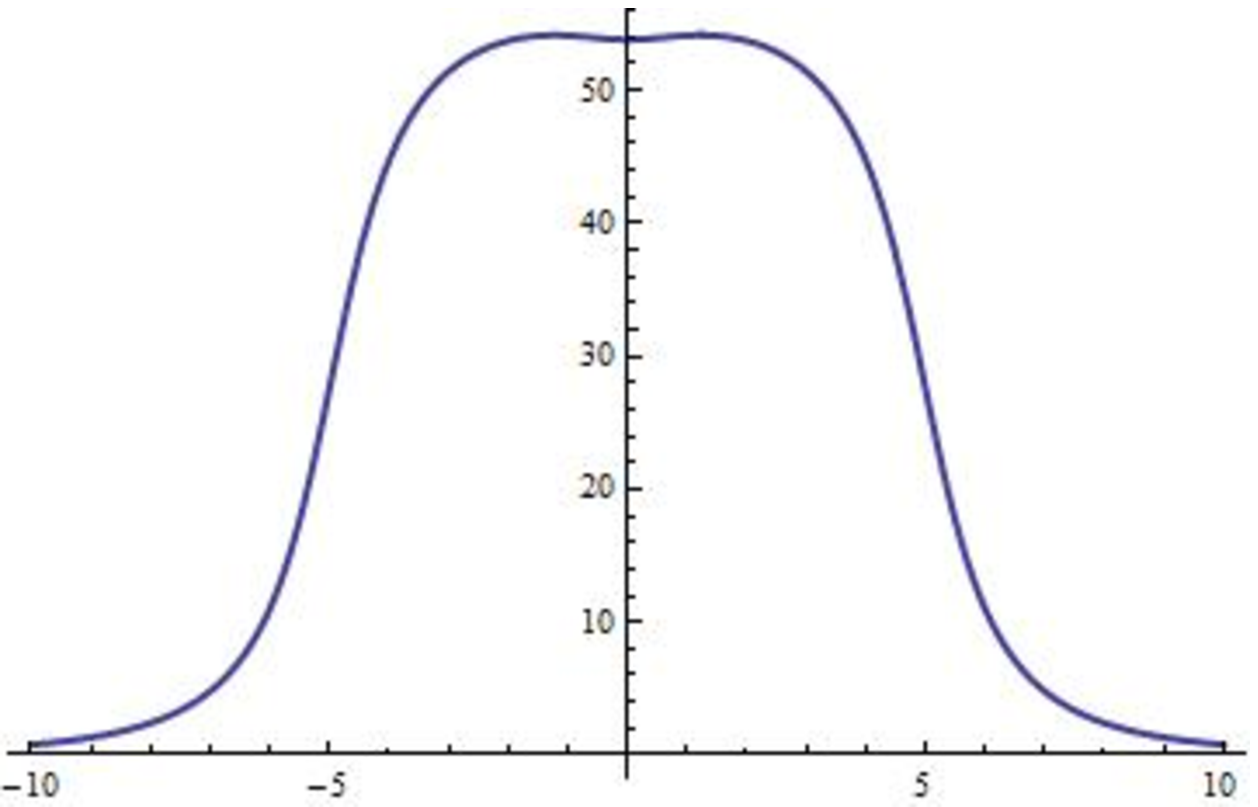}
   \end{center}
   \caption{The exact plot of $\Delta S_A$ as a function of $t$ for $(l^{(1)},l^{(2)})=(0,5)$.
   The left, middle and right graph correspond to $M=0.01,5,500$, respectively. We set $R=\ap=4G_N=1$.
    }\label{fig:timedeg}
\end{figure}

Especially, let us focus on the case $(l^{(1)},l^{(2)})=(0,l)$. In
this case, one of the end points of $A$ i.e. $x=0$ is excited by the
local quench. The result is plotted in Fig.\ref{fig:timedeg}. It is
easy to see that $\Delta S_A$ is non-trivial during $-l\lesssim
t\lesssim l$, as expected from the causality argument. This
qualitatively agrees with the CFT result in \cite{CaCaL}. As $M$
gets larger, the HEE looks like a step function.

In the late time limit $t\to \infty$, $S_A$ approaches to the result of the ground state
\be
S_A(t=\infty)=\f{c}{3}\log \f{l}{z_{\infty}}. \label{eegr}
\ee
On the other hand, at the time $t=0$, we find the following result for $l>>\ap$
\be
S_A(t=0)=S_A(t=\infty)+\f{c}{6}
\log\left(\f{R^2}{R^2-M}\cdot\sin^2\left(\f{\pi\s{R^2-M}}{2R}\right)\right).
\ee
Notice that we always have  $S_A(t=0)\geq S_A(t=\infty)$.

\subsubsection{Semi Infinite Limit $l\to \infty$}

\begin{figure}[ttt]
   \begin{center}
     \includegraphics[height=3cm]{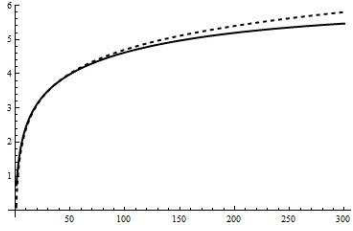}
   \end{center}
   \caption{The approximation of $\Delta S_A$ by the function $\log t+$const.
   for $(l^{(1)},l^{(2)})=(0,1000)$. The thick black curve is the
   plot of the exact result, while the dashed one is its
   approximation by $\log t+0.1$. We choose $M=0.5$ and set
   $R=\ap=4G_N=1$. They nicely agree with each other when $\ap<<t<<l$.
    }\label{fig:timedlog}
\end{figure}

It is useful to consider the limit $l>>t>>\ap$ for $(l^{(1)},l^{(2)})=(0,l)$. We find
\ba
&& \tau^{(1)}_{\infty}\simeq \pi-\f{2\ap}{t},\ \ \
\tau^{(2)}_{\infty}\simeq \f{2t\ap}{l^2},\no
&& \theta^{(1)}_\infty= 0,\ \ \ \theta^{(2)}_\infty\simeq \pi-\f{2\ap}{l},\no
&& r^{(1)}_{\infty}\simeq \f{Rt^2}{2\ap z_{\infty}},\ \ \
r^{(2)}_{\infty}\simeq \f{Rl^2}{2\ap z_{\infty}}.
\ea
Finally the HEE is found to be
\ba
&& S_A\simeq \f{c}{3}\log \f{Rtl}{2\ap z_{\infty}}+\f{c}{6}\log \left[\f{4}{R^2-M}\cdot
\sin\left(\f{\s{R^2-M}}{R}\left(\pi-\ap/t\right)\right)\cdot
\sin\left(\f{\ap\s{R^2-M}}{Rt}\right)\right]\no
&& \ \ \  \simeq \f{c}{3}\log \f{l}{z_{\infty}}+\f{c}{6} \log \f{t}{\ap }
+\f{c}{6}\log\left(\f{R}{\s{R^2-M}}\sin\left(\f{\pi\s{R^2-M}}{R}\right)\right).
\ea
In this way, we obtain the behavior
\be
S^{AdS}_A\sim \f{c}{6}\log \f{t}{\ap}+\f{c}{3}\log \f{l}{z_{\infty}}, \label{btlo}
\ee for the time evolution. Refer to
Fig.\ref{fig:timedlog} for the numerical confirmation.

On the other hand, the local quench induced by joining two half lines leads to the following behavior \cite{CaCaL}
\be
S^{joint}_A\sim \f{c}{3}\log \f{t}{z_{\infty}}+\f{c}{6}\log\f{l}{\ep}, \label{btloo}
\ee
where $\ep$ is the cutoff for the process of local quench introduced in \cite{CaCaL} and is analogous to $\ap$ in our model.

Note that both have the common important property that $S_A$
increases
 logarithmically after local quenches. This is a characteristic feature of local quenches in two dimension. To be more precise, the coefficient of $\log t$ is different
 between (\ref{btlo}) and (\ref{btloo}). Indeed, it is known that
 this coefficient depends on how we locally excite the system as found in the example analyzed in \cite{Eisler}. The difference of the coefficient in front of $\log l$ is easy to understands because in the case (\ref{btlo}) the system was already joined before the quench, while in the other case (\ref{btloo}), the system was originally disconnected.
 We will give a more detailed explanation of the two difference behaviors (\ref{btlo})
and (\ref{btloo}) in section 6.2 using the tensor network description.

\subsection{HEE for General Intervals}

\begin{figure}[ttt]
   \begin{center}
     \includegraphics[height=3cm]{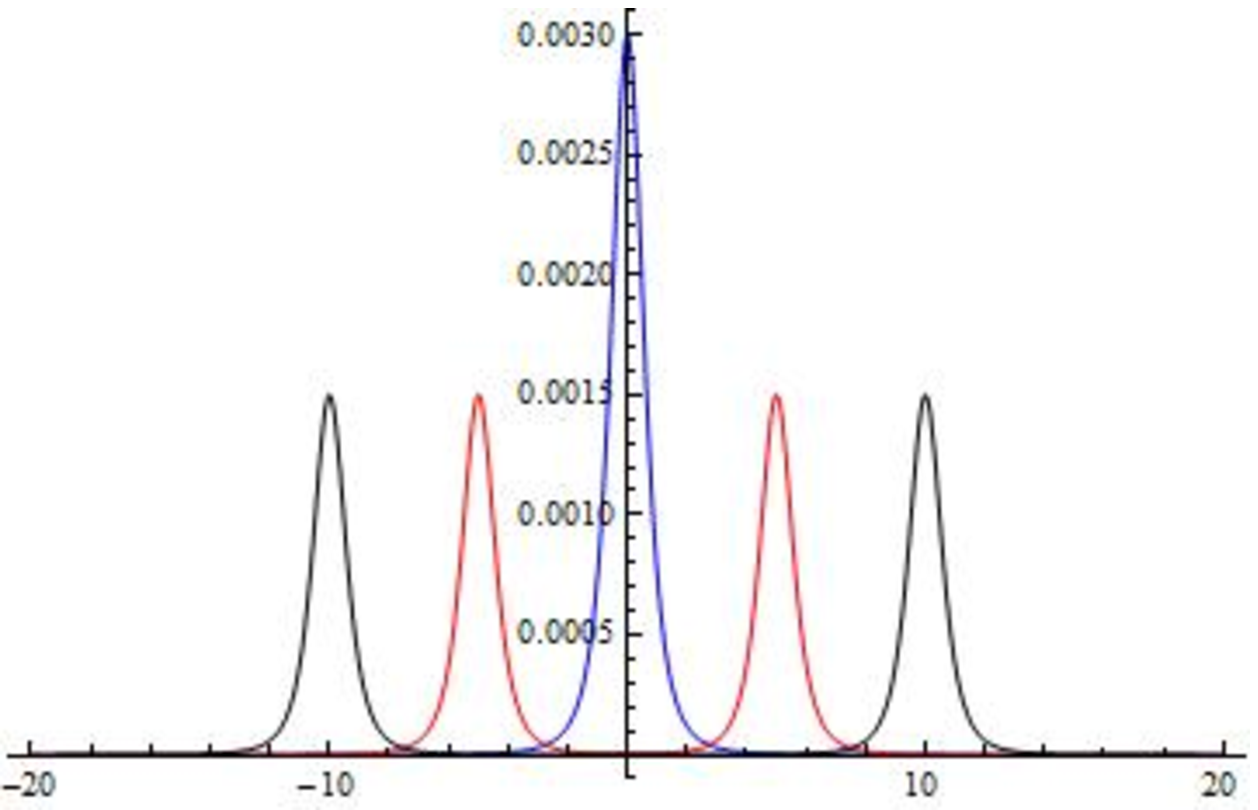}
     \includegraphics[height=3cm]{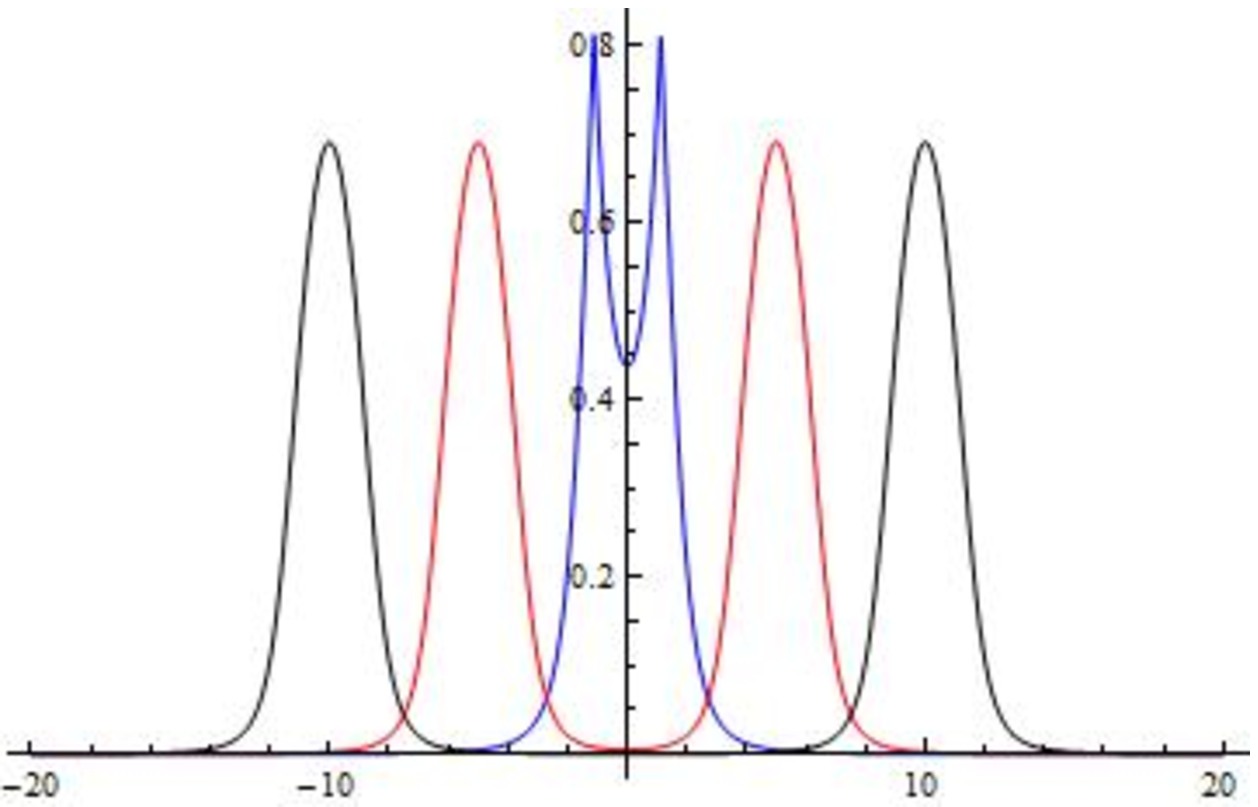}
     \includegraphics[height=3cm]{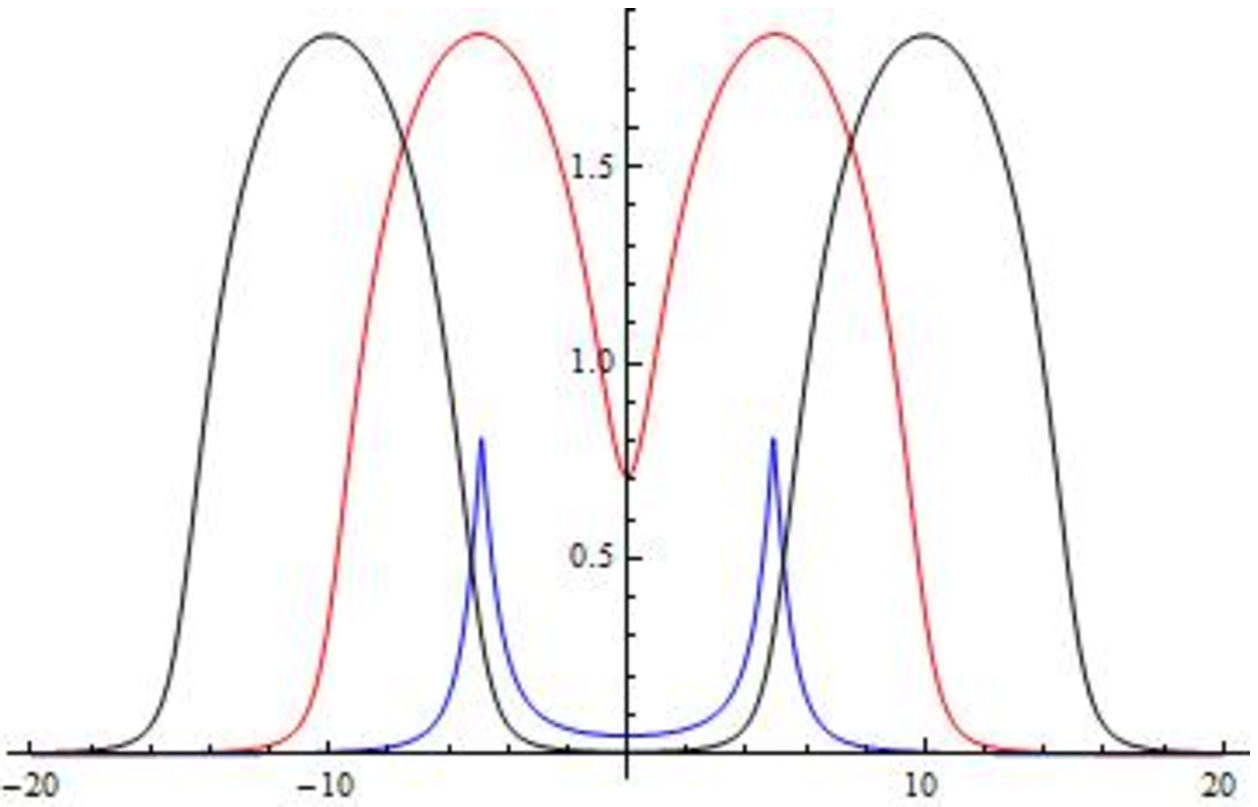}
   \end{center}
   \caption{The plot of $\Delta S_A$ as a function of $\xi$. We set $l=0.1$ (left), $l=3$ (middle) and $l=10$ (right). In each graph, the blue, red and black curve describes $\Delta S_A(l,\xi,t)$
   for $t=0$, $t=5$ and $t=10$, respectively. We choose $M=0.9$ and set $R=\ap=4G_N=1$.}\label{fig:SAxi}
\end{figure}

Now we would like to turn to the HEE $S_A$ for a subsystem
$A$ defined by a general interval $(l^{(1)},l^{(2)})$.
Equally we can think that $A$ is parameterized by
its width $l$ and the position of its center $\xi$ via
\be
l^{(1)}=\xi-l/2,\ \ \ l^{(2)}=\xi+l/2. \label{lxi}
\ee
We can explicitly calculate $S_A(l,\xi,t)$ by using the formula
(\ref{genints}). We plotted $\Delta S_A$ as a function of $\xi$
for fixed values of $l$ and $t$ in Fig.\ref{fig:SAxi}.

First, we notice that for a small $l$ (the left graph of
Fig.\ref{fig:SAxi}), the initial peak at $\xi=0$ divides into two
peaks under time evolution, keeping total sum of the heights
conserved. This result looks very similar to the energy density
$T_{tt}$ in two dimensional CFTs (see the left graph in
Fig.\ref{fig:EMtime}), where the energy conservation holds. Indeed,
we can easily understand this coincidence from the first law
relation (\ref{firstEE}) in the small $l$ limit, which can be
rewritten as \be \Delta S_A(l,\xi,t)\simeq \f{\pi
l^2}{3}T_{tt}(\xi,t). \label{saen} \ee Therefore we can conclude
that in the small $l$ limit, the integral \be
\int^{\infty}_{-\infty} d\xi~ \Delta S_A(l,\xi,t), \label{consee}
\ee does not depend on the time $t$.

\begin{figure}[ttt]
   \begin{center}
     \includegraphics[height=3cm]{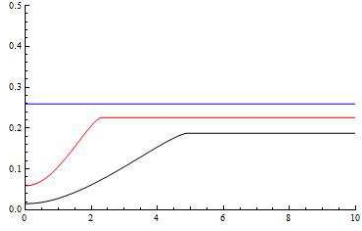}
   \end{center}
   \caption{The numerical plot of $\f{1}{l^2}\int d\xi \Delta S_A(l,\xi,t)$  as a function of time $t>0$.
   The blue, red and black graph correspond to $l=1,5,10$, respectively. In the small $l$ limit, this quantity
   should approach $\f{\pi}{12}\simeq 0.262$ as follows from (\ref{saen}).
   We choose $M=0.5$ and set $R=\ap=4G_N=1$.
    }\label{fig:sum}
\end{figure}

Moreover, for $l>\ap$, we find that $\Delta S_A$ at $t=0$ has a peak
at $|\xi|=l/2$. In the gravity dual, this is easy to understand
because the geodesic $\gamma_A$ comes very close to the massive
particle as explained in Fig.\ref{fig:setup}. At a later time
$t>l/2$, the hump propagates at the speed of light, conserving its
form. This occurs because $\gamma_A$ is now far from the falling
particle and its shock waves propagates without changing its form.
Therefore we expect that the integral (\ref{consee}) will approach
to a constant value at late time $t>l/2$. Indeed, we can confirm
this numerically as shown in the Fig.\ref{fig:sum}. This behavior
roughly tells us that the long range entanglement will be generated
at late time, while the short range one will be not. We will study
more carefully such a dynamical structure of quantum entanglement in
the next section, which will provide clear explanations for all of
the results in this section.

\section{Entanglement Density}

In this section, we will introduce a new quantity which we call
entanglement density. It can extract an essential structure of
quantum entanglement from $S_A$ in two dimensional quantum field
theories. We will apply this to our holographic local quenches in
AdS$_3/$CFT$_2$.

\subsection{Definition of Entanglement Density}

Consider a general two dimensional field theory in a certain pure state and suppose that we can calculate its entanglement entropy $S_A$. The subsystem $A$ is defined by the interval $l^{(1)}\leq x\leq l^{(2)}$ (or more simply $[l^{(1)},l^{(2)}]$), where $x$ is the space coordinate of the field theory. Just for convenience, we will treat $l^{(1)}$ and $l^{(2)}$ to be independent parameters for a while. We would like to ask how to extract a structure of quantum entanglement explicitly from the entanglement entropy $S_A$ for arbitrary intervals.

For this purpose, we take into account only the two body entanglement and define the {\it entanglement density} $n(l,\xi,t)$ such that this counts the number of entangling pairs (or bits) between the two points $x=\xi-l/2(=l^{(1)})$ and $x=\xi+l/2(=l^{(2)})$. Imagine that we discretize the field theory into a lattice theory such as spin chains. In each point, we assume that there are several spins. Then $n(l,\xi,t)$ counts the number of entangled pairs of spins which are located at $x=\xi-l/2$ and $x=\xi+l/2$. Therefore the parameter $l$ describes the range of the entanglement and $\xi$ does the position of the center of the entangled pair.

\begin{figure}[ttt]
   \begin{center}
     \includegraphics[height=6cm]{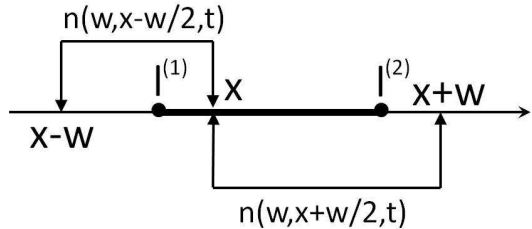}
   \end{center}
   \caption{A sketch of expression of $S_A$ as a sum of entanglement density.
   Hooks with the arrows on both end points describe the entangling pairs between two points.}\label{fig:density}
\end{figure}

Now we approximate $S_A$ by summing all entangling pair as follows (see Fig.\ref{fig:density}):
\ba
S_A=\int^{l^{(2)}}_{l^{(1)}} dx\left[\int^\infty_{x-l^{(1)}}dw~n(w,x-w/2,t)+\int^\infty_{l^{(2)}-x}dw~n(w,x+w/2,t)\right].
\ea
By taking a derivative with respect to $l^{(2)}$ with $l^{(1)}$ fixed, we obtain
\ba
\f{\de S_A}{\de l^{(2)}}&=&\int^\infty_{l^{(2)}-l^{(1)}}
dw~n\left(w,l^{(2)}-\f{w}{2},t\right)+\int^\infty_{0}dw~n\left(w,l^{(2)}+\f{w}{2},t\right) \no
&& -\int^{l^{(2)}}_{l^{(1)}} dx~ n\left(l^{(2)}-x,\f{x}{2}+\f{l^{(2)}}{2},t\right).
\ea
 Finally by taking the derivative by $l^{(1)}$, we find
 \be
 \f{\de^2 S_A}{\de l^{(1)}\de l^{(2)}}=\f{1}{4}\f{\de^2 S_A}{\de \xi^2}-\f{\de^2 S_A}{\de l^2}=2n\left(l^{(2)}-l^{(1)},\f{l^{(1)}+l^{(2)}}{2},t\right)=2n(l,\xi,t). \label{defentd}
 \ee
 In this way, we can extract $n(l,\xi,t)$ if we know $S_A$ as a function of $l^{(1)},l^{(2)}$ and $t$.

For a ground state of a CFT$_2$ with the central charge $c$,  $S_A$
is given by (\ref{eegr}) and thus the entanglement density is found
to be \be n_{CFT}(l,\xi,t)=\f{c}{6l^2}. \label{Ncft} \ee Notice that
the entanglement density is free from the UV divergences. Thus, when
we talk about a deformation of a CFT$_2$ as in the present example,
it is sometimes useful to normalize the entanglement density by
multiplying $l^2$.

\subsection{Strong Subadditivity and Positivity of Entanglement Density}

Consider a two dimensional field theory in a generically excited
state. We introduce three subsystems $A_1,A_2$ and $A_3$ so that
they are defined by the intervals $[l_1,l_2],[l_2,l_3]$ and
$[l_3,l_4]$, respectively, such that $l_1<l_2<l_3<l_4$.

The strong subadditivity \cite{LR} (refer to \cite{CHS} for its
application to field theories and to \cite{HT} for its holographic
proof) is given by the inequality \be S_{A_1\cup A_2}+S_{A_2\cup
A_3}\geq S_{A_1\cup A_2\cup A_3}+S_{A_2}. \label{SSA} \ee In our setup, this is
equivalent to \be S(l_1,l_3)+S(l_2,l_4)\geq S(l_2,l_3)+S(l_1,l_4),
\ee where $S(p,q)$ is the entanglement entropy $S_A$ for the
interval $[p,q]$.

In particular, we can take the limit $\delta_{1,2}\to +0$ with
$l_2=l_1+\delta_1$ and $l_4=l_3+\delta_2$. Then, the strong
subadditivity (\ref{SSA}) is rewritten into \be \f{\de^2 S(p,q)}{\de p\de q}\geq
0. \ee This proves that the entanglement entropy $n(l,\xi,t)$ in
(\ref{defentd}) is positive.

\subsection{Conservation Law}

It is useful to study how the total integration of the entanglement
density behaves. To make the analysis simple, we assume that the $x$
coordinate can be compactified on a circle with the length $L$. The
infinitely extended system we are mainly studying in this paper can
be obtained in the limit $L\to \infty$. Since we have the obvious
relation $n(l,\xi,t)=n(L-l,L-\xi,t)$ due to the periodicity, we can
restrict the integration of $l$ to $0\leq l \leq L/2$. We will focus
on the increased amount of entanglement entropy and entanglement
density compared with the ground state, denoted by $\Delta S_A$ and
$\Delta n$, respectively.

Thus we find \ba && \int^{L}_{0} d\xi \int^{L/2}_0 dl~ \Delta
n(l,\xi,t)= \int^L_0 dl^{(1)}\int^{l^{(1)}+L/2}_{l^{(1)}} dl^{(2)} ~
\Delta n(l,\xi,t) \no && = \f{1}{2}\int^L_0
dl^{(1)}\int^{l^{(1)}+L/2}_{l^{(1)}} dl^{(2)} \f{\de^2 \Delta
S_A}{\de l^{(1)}\de l^{(2)}} \no && = \f{1}{2} \int^{L}_0
dl^{(1)}\left[\f{\de}{\de x}\Delta S_A(x,l^{(1)}+L/2)\Bigl
|_{x=l^{(1)}}-\f{\de}{\de x}\Delta S_A(x,l^{(1)})\Bigl |_{x=l^{(1)}}
\right].\label{consden} \ea We can show that the first term in the
third line of (\ref{consden}) vanishes. This is because by employing the
identity $S_A=S_B$ for pure states and the periodicity of the
circle, we can easily show \be \int^L_0 dy \left[\Delta
S_A(y+\delta/2,y+L/2)-\Delta S_A(y-\delta/2,y+L/2)\right]=0, \ee for
any $\delta$. Moreover, the other term in the third line of
(\ref{consden}) vanishes because the first law (\ref{saen}) tells
us \be \Delta S_A(y+\delta/2,y)-\Delta S_A(y-\delta/2,y)\simeq
\f{\pi}{3}T_{tt}\cdot \delta^2, \ee in the $\delta \to 0$ limit.
Remember that our first law assumes that there is a UV fixed point
in the field theory we consider.

In this way, we proved that the total integration of entanglement
density is constant. In other words, the total number of entangled
pairs is conserved: \be \int^{L}_{0} d\xi \int^{L/2}_0 dl~ \Delta
n(l,\xi,t)=0. \label{necons} \ee

In general, the negative contribution comes from the UV region where
$l$ is very small. Indeed, by using the first law (\ref{saen}), we
can generally show the following result in the limit $l\to 0$: \be
\lim_{l\to 0}\Delta n(l,\xi,t)=-\f{\pi}{3}T_{tt}(\xi,t).
\label{negct} \ee

\subsection{Entanglement Density for Global Quenches}

To understand better the meaning of entanglement density, we would like to consider the global quenches
in two dimensional CFTs with central charge $c$ before we examine the local quenches. As shown in \cite{CaCaG}, the increased amount of entanglement entropy for the subsystem with the width $l$ behaves like
\be
\Delta S_A=c\cdot \Delta m\cdot t\ \ (\ep<t\leq l/2), \ \ \ \ \Delta S_A=\f{c}{2}\cdot \Delta m\cdot l\ \ (t\geq l/2),
\ee
where $\Delta m$ is the energy gap which creates the global quench times a numerical constant and is
proportional to the effective temperature at late time; $\ep$ is cut off scale\footnote{
 More precisely this is the length scale beyond which we can approximate the excited state by the boundary state, which preserves the boundary conformal invariance. See \cite{TaUg} for more details.} of the quench process, which is of order $\delta m^{-1}$.

\begin{figure}[ttt]
   \begin{center}
     \includegraphics[height=6cm]{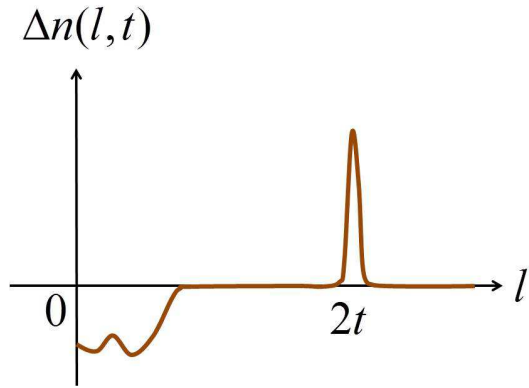}
   \end{center}
   \caption{A sketch of the entanglement density as a function of $l$ under a global quench. }\label{fig:globalQ}
\end{figure}

In this case, the entanglement density (\ref{defentd}) reads
\be
\Delta n(l,\xi,t)=\f{c\cdot \Delta m}{4}\cdot \delta(t-l/2)+ \Delta n(l,\xi,t)_{UV}, \label{nfuncl}
\ee
where the last term denotes the UV contribution which is non-trivial only for $l$ of order $\ep\sim \Delta m^{-1}$ or smaller. This UV term is negative such that the conservation law (\ref{necons}) holds. The delta functional term in (\ref{nfuncl}) shows that the entangled pairs have the range $2t$ at time $t$. We sketched this behavior in Fig.\ref{fig:globalQ}. This behavior can be understood that the
entangled pair moves in the opposite way at the speed of light, being consistent with the interpretation in \cite{CaCaG}.

In this way, for the excited states produced by quantum quenches,
$\Delta n(l,\xi,t)$ tends to positive for large $l$, while it
becomes negative for small $l$ due to the conservation law and the
first law as in (\ref{negct}). Therefore we learn that an excitation
of a quantum system corresponds to breaking shorter range
entanglement and creating longer range entanglement.

\subsection{Entanglement Density for Holographic Local Quenches}

\begin{figure}[ttt]
   \begin{center}
     \includegraphics[height=3cm]{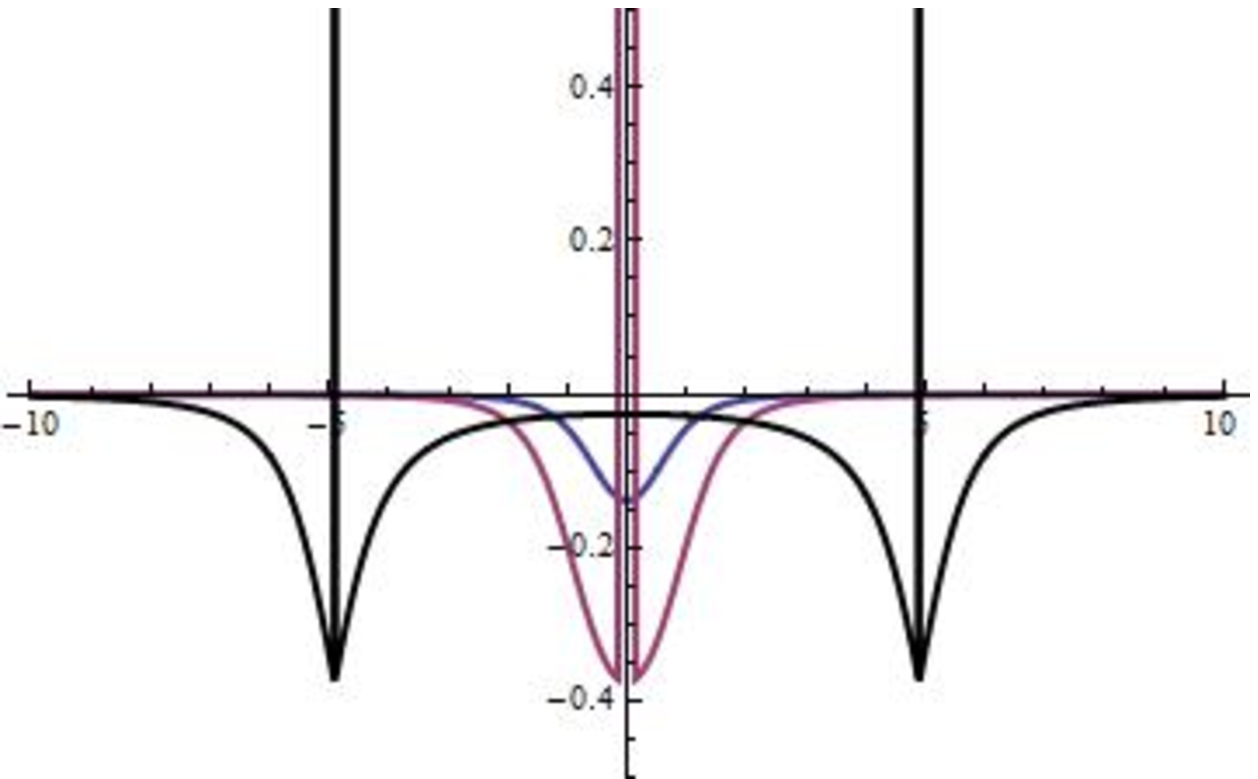}
     \hspace{5mm}
       \includegraphics[height=3cm]{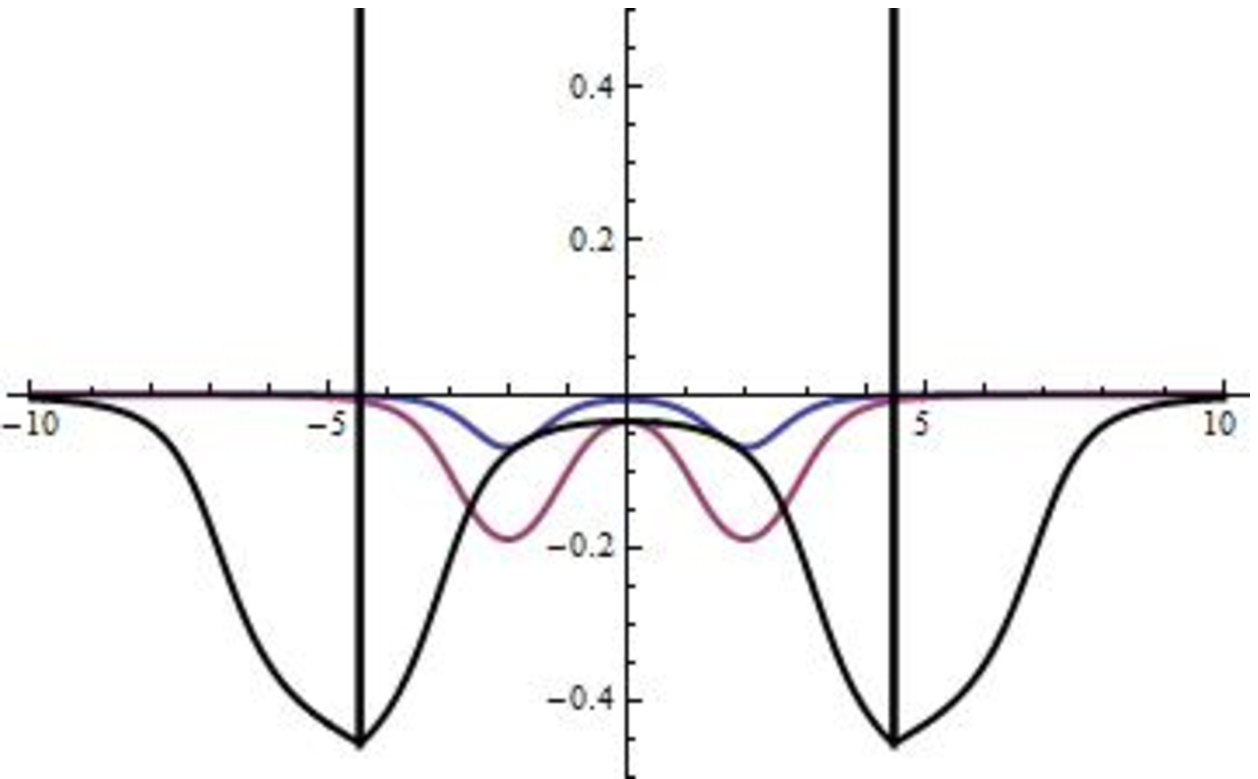}
   \end{center}
   \caption{The numerical plots of $l^2 \Delta n(l,\xi,t)$  as a function of the center position $\xi$ at $t=0$ (left graph) and $t=2$ (right graph).
   The blue, red and black graph correspond to $l=1,2,10$, respectively.
   We choose $M=0.5$ and set $R=\ap=4G_N=1$. Notice the relation $l^2n(l,\xi,t)=1+l^2\Delta n(l,\xi,t)$
    }\label{fig:nent}
\end{figure}

Now let us come back to the analysis of holographic local quenches. We consider the difference from the ground state and would like to calculate $\Delta n(l,\xi,t)$ from $\Delta S_A(l,\xi,t)$ by using the formula
(\ref{defentd}). The results are plotted in Fig.\ref{fig:nent}. We can confirm that the total (properly normalized) density $l^2 n(l,\xi,t)=1+l^2\Delta n(l,\xi,t)$ is always positive as expected (we set $c=6$ in the plot).

First notice that the positive peaks in Fig.\ref{fig:nent} are all delta-functional. This is because our gravity backgrounds, where we calculated the HEE, are singular due the presence of the deficit angle, corresponding to the point particle limit. This delta-functional peak appears at
\be
\xi=\pm\s{l^2/4-t^2-\ap^2},  \label{curvecon}
\ee
which is when the massive particle is on top of $\gamma_A$ (see Fig.\ref{fig:setup}).
However, it is easy to imagine a regularization by replacing the point particle with a finite size object, which will replace a delta-functional peak with a smooth peaks with a finite width (called $\eta$).

\begin{figure}[ttt]
   \begin{center}
     \includegraphics[height=3cm]{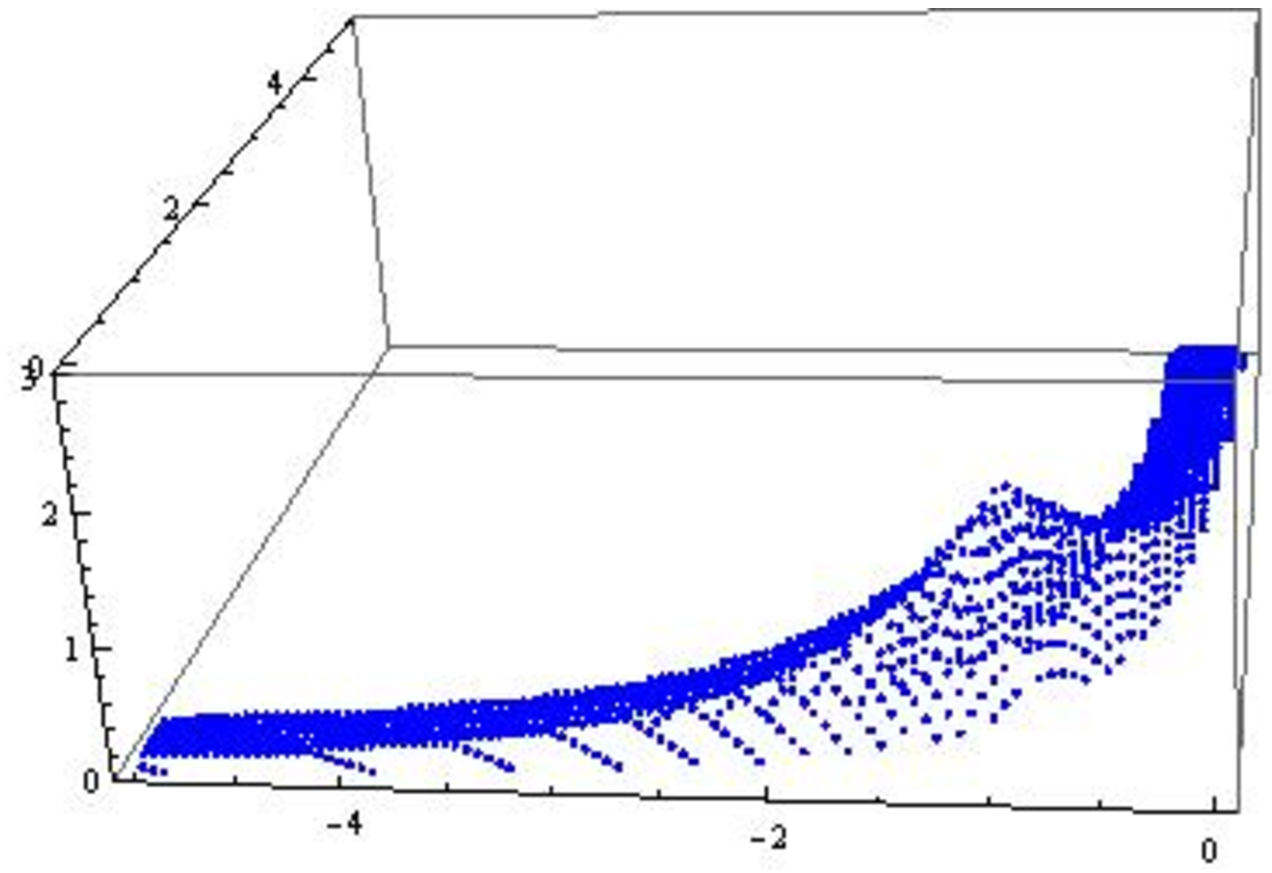}
     \hspace{5mm}
       \includegraphics[height=3cm]{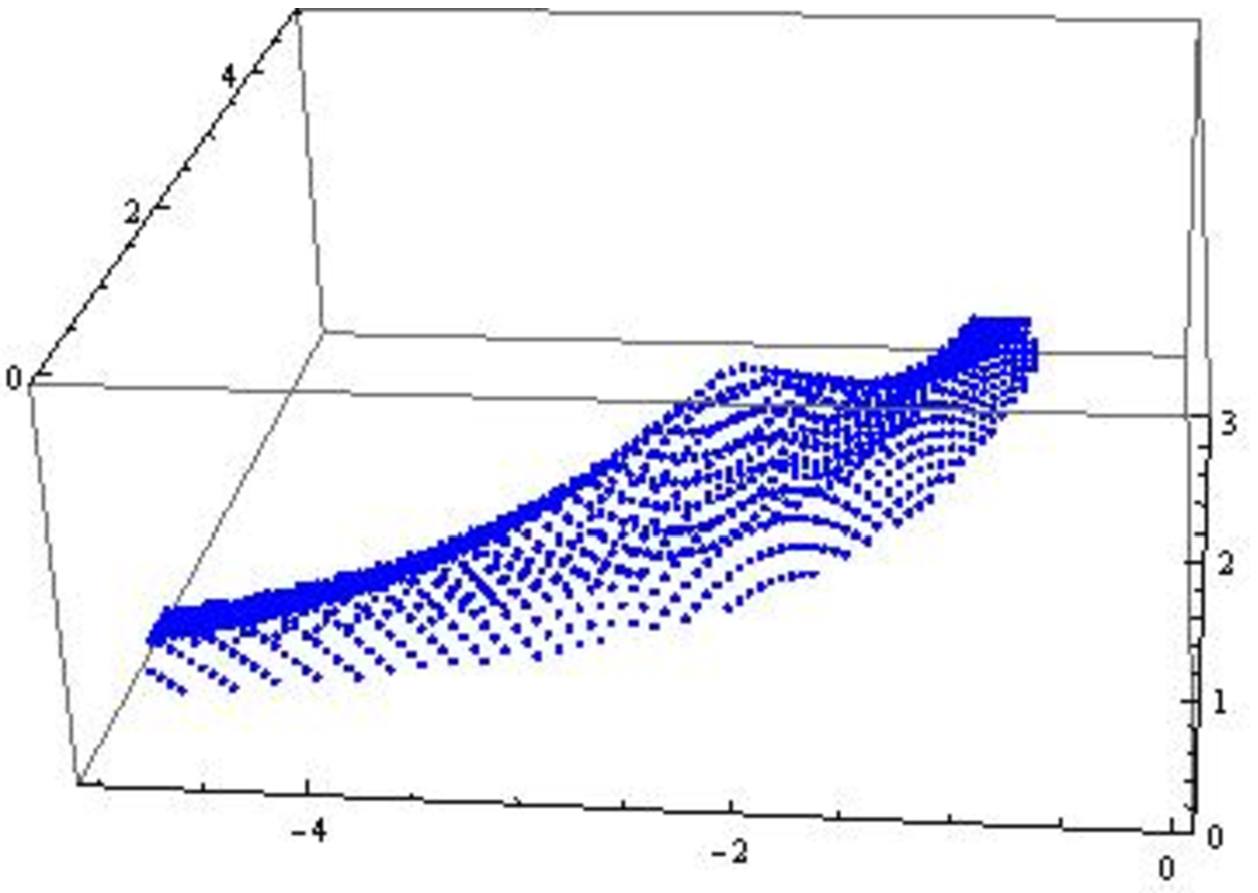}
   \end{center}
   \caption{The numerical plots of $\Delta n(l,\xi,t)$ as a function of $l^{(1)}=\xi-l/2<0$ (horizontal axis) and
   $l^{(2)}=\xi+l/2>0$ (inward axis). We choose the regularization parameter as $\eta=0.4$. The left and right plot
   corresponds to $t=0$ and $t=2$, respectively. We only plot those points where $\Delta n(l,\xi,t)$ is positive.   We choose $M=0.5$ and set $R=\ap=4G_N=1$.}\label{fig:nentev}
\end{figure}

Employing this regularization,\footnote{More explicitly, we replaced
the derivative $\f{\de^2 S_A}{\de l^{(1)}\de l^{(2)}}$ with the second order difference $S_A(l^{(1)}+\eta/2,l^{(2)}+\eta)-S_A(l^{(1)}-\eta/2,l^{(2)}+\eta)
-S_A(l^{(1)}+\eta/2,l^{(2)}-\eta)+S_A(l^{(1)}-\eta/2,l^{(2)}-\eta)$.} the
entanglement density $n(l,\xi,t)=n(l^{(2)}-l^{(1)},(l^{(2)}+l^{(1)})/2,t)$
is plotted as a function of $l^{(1)}$ and $l^{(2)}$ in Fig.\ref{fig:nentev} for $t=0$ and $t=2$.
It is clear from this graph that the peaks exist along the curve (\ref{curvecon}) or equally
$l^{(1)}\cdot l^{(2)}=-\ap^2-t^2$ as expected from the condition that the massive particle passes through $\gamma_A$. Moreover, $\Delta n$ takes the largest value when $l^{(2)}=-l^{(1)}=\s{\ap^2+t^2}$.

\begin{figure}[ttt]
   \begin{center}
     \includegraphics[height=6cm]{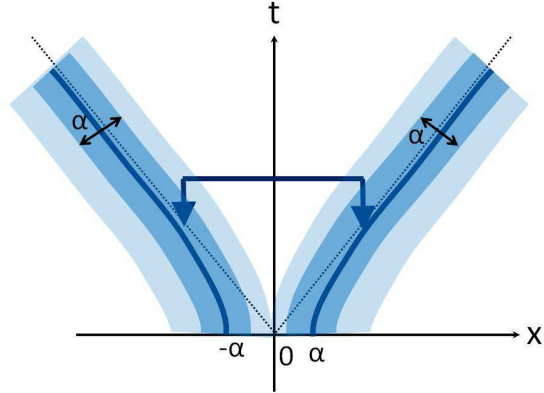}
   \end{center}
   \caption{A sketch of time evolution of entangling pairs. The blue thick curve describes the dominant
   entangled pair $\xi=0$. Around this, other entangled pairs are gathered with the width of order $\ap$.
   }\label{fig:evolution}
\end{figure}

These behaviors can be
systematically understood by considering the time-evolution of each entangled pair\footnote{Another holographic realization of entangled pairs refer to \cite{FTT} based on the AdS/BCFT proposal \cite{Ta}.}. At time $t$, the pair of the points $x=\xi-\s{\xi^2+\ap^2+t^2}$ and  $x=\xi+\s{\xi^2+\ap^2+t^2}$ are entangled for any values of $\xi$. Especially the pair at $\xi=0$ (i.e. the one between $x=-\s{\ap^2+t^2}$ and
$x=\s{\ap^2+t^2}$) possesses the strongest entanglement, which we call the dominant entangled pair.
More generally the entanglement is enough strong for the entangled pairs with $|\xi|\lesssim \ap$.
Therefore they follow the time evolutions in an obvious way, as sketched in Fig.\ref{fig:evolution}. At earlier time $t<<\xi$ the pair moves slowly. At late time $t>>\xi$, they move at the speed of light in the opposite direction. This time evolution is also sketched with the energy density in Fig.\ref{fig:pair}.

We can explain the qualitative behaviors of HEE calculated in the previous section by referring to
this structure of quantum entanglement in our holographic local quenches, based on the entangled pairs.
For example, let us explain the result described in Fig.\ref{fig:sum}.
 First assume $t<l/2$. Then if we
sweep a length $l$ interval $A$ by shifting its center $\xi$,  $S_A$ cannot detect the entanglement pairs
generated by the local quench when $t-l/2\leq \xi\leq -t+l/2$ because the dominant entangled pair is
completely included in $A$ for these values of $\xi$. This missing contribution is proportional to
$l/2-t$ and therefore $S_A$ increases linearly until $t=l/2$.
At late time $t>l/2$, for any
$\xi$, the interval $A$ always include only one of the pair and therefore it fully contributes to
$S_A$. Thus the integral $\int d\xi \Delta S_A$ no longer depends on the time $t$. These explain the behavior found in Fig.\ref{fig:sum}.

One may think that any long range entanglement should not exist at $t=0$ because the massive particle is situated at deep UV region $z=\ap$. However, this speculation is based on a too naive UV/IR relation of AdS/CFT. Indeed, if we consider a two point function $\la O(x)O(y)\lb$ for a certain operator with a large conformal dimension, then we can evaluate this from the geodesic distance in AdS. The same geodesic distance appears in the calculation of HEE and therefore we find that a non-trivial result for two point functions can obtained when $x=0$, where the geodesic $\gamma_A$ passes very close to the massive particle as in right picture of Fig.\ref{fig:setup}. Thus this consideration of two point functions also support the long range entanglement
even at $t=0$.  Another way to see this is to note that the energy density $T_{tt}$ computed
in (\ref{Ttt}) decays only slowly as $x^{-4}$ in the long distance limit $x\to \infty$
for any non-zero $\ap$.

In this way, we learned that the long range
entanglement is non-trivial even at $t=0$ in our holographic model. In ideal models of local quenches we may not
expect any long range entanglement just after the quench. However this is not a major problem because as we mentioned the dominant entangled pair is the one between $x=-\s{\ap^2+t^2}$ and
$x=\s{\ap^2+t^2}$, which is short range ($\sim \ap$) at $t=0$.

In summary, we can understand the time evolutions of entanglement
entropy $S_A$ in terms of those of entanglement density $n$.  In our
holographic local quench, the dynamical behavior of $n$ is described
by the evolution of the entangled pairs as in
Fig.\ref{fig:evolution}. At late time, the range of entanglement
increases at the speed of light. As the time evolves, short range
entanglement disappears and long range one is generated so that the
total number of entangled pairs is conserved as in (\ref{necons}).
This process looks a sort of the decoherence phenomena.

\section{Quantum Information, Thermodynamics and Gravity}

In this section we employ the results found in the previous sections
to develop some insights on connections between the quantum
information and gravity via the AdS/CFT correspondence.

\subsection{Total Amount of Quantum Information}

In the CFT side, the entanglement entropy $S_A$ is interpreted as
the entropy for an observer in $A$, assuming that the observer is
not accessible to the region $B$. In order words this measures the
amount of quantum information inside $B$ for such an observer.
Therefore $\Delta S_A$ can be regarded as the amount of quantum
information in $B$ which is induced by the local quench, such as the
data of excited particles etc.

Notice that the standard thermal entropy for a mixed state, which is
dual to a black hole entropy via AdS/CFT, corresponds to the
information which is not accessible to the observer by any means
(except for making microscopic observations of its heat bath). On
the other hand, the entanglement entropy $S_A$ measures the amount
of information which is possessed by $B$ and is accessible to the
observe in $A$ after all possible experiments are done.

Therefore, the holographic entanglement entropy allows us to
calculate such a novel quantity, i.e. the amount of information
$\Delta S_A$, for any massive object in AdS, which is dual to a
certain excited state in the CFT. Apply this idea to our falling
particle in AdS. The standard principle in AdS/CFT tells us that the
size $l$ of the localized excitations induced by the local quench is
estimated as $l\sim \s{t^2+\ap^2}$, which is the value of the
coordinate $z$ of the falling particle. Therefore it is natural that
we choose the subsystem $A$ (or equally\footnote{Notice that since
we are considering a pure state we always have $S_A=S_B$ and
therefore the distinction between $A$ and $B$ is not important.}
$B$) to be the ball with the radius $\s{t^2+\ap^2}$. Indeed, we can
see from our perturbative results (\ref{parea4}) and (\ref{peet})
that $\Delta S_A$ takes the maximal value when $l= \s{t^2+\ap^2}$ as
a function of $l$ and takes a time-independent constant value.
Therefore we call it $\Delta S^{max}$. This is evaluated as the
following simple form for any dimension $d$: \be \Delta
S^{\max}=C_s\cdot mR=C_s \cdot \Delta, \ee where $C_s$ is an order
one constant. Explicitly, in each dimension $d$, we find $C_s=2$
($d=2$), $C_s=\pi/2$ ($d=3$) and $C_S=4$ ($d=4$). We used the
well-known relation between the mass $m$ of a particle and the
conformal dimension $\Delta$ of its dual operator \cite{GKPW}
assuming $\Delta >>1$. For example, in the AdS$_3$ case the exact
calculation (\ref{finco}) leads to at $l=\s{t^2+\ap^2}$: \be \Delta
S^{max}=\f{c}{3}\log \left(\f{R}{\s{R^2-M}}
\sin\left(\f{\pi\s{R^2-M}}{2R}\right)\right). \label{sam} \ee This
shows that $\Delta S^{max}$ is a constant which is independent from
$l$. Assuming the range $1<<\Delta <<c$ we find\footnote{If we
assume a very large dim. operator $\Delta >>c$ in the local thermal
quench, then it is dual to a BTZ black hole. In this case we find
from (\ref{sam}) $\Delta S^{max}\simeq \f{\pi\s{M}}{4G_N}=
\f{\pi}{3}\s{3c\Delta}$. This is a half of the black hole entropy
and agrees with the Cardy formula.} \be \Delta S^{max}\simeq
2\Delta. \ee

In this way we find the number of miscrostates of the
massive particle is estimated as follows
in any $d$:
\be
\#~ \mbox{Microstates} \sim e^{\Delta S^{max}}= e^{C_s\cdot \Delta}. \label{micro}
\ee

This is qualitatively consistent with the well-known Hagedron
growth of degeneracy of the states with a large conformal
dimension $\Delta >>1$ \cite{Witten,Hage}.
Remember that this relation (\ref{micro})
holds only when the backreaction of the massive
particle is small. For example, this requires $\Delta <<c$
in $d=2$ as can be seen from
(\ref{sam}).

If we remember that the energy $E$ of localized excitations made of
radiations in the CFT is given by (\ref{energy}), then we find the
simple relation between the amount of information of this excited
lump and its the total energy:
\be
\Delta S^{max}\sim E\cdot \ap. \label{ienergy}
\ee
Remember that $\ap$ denotes the size of the lump at the
initial time $t=0$. This provides us a prediction that in the
large $N$ strong coupled gauge theories,
the amount of information possessed by such a radiation `fire ball'
is given by its energy $E$ times its linear size $\ap$.
It would be very intriguing future problem to extend this
estimation to other objects like the fundamental strings or
D-branes in AdS/CFT in order to see how much the relation
(\ref{ienergy}) is universal.

 Finally, we would also like to comment on a possible
 interpretation of $\Delta S^{max}$ from the gravity side.
 Since the Unruh temperature of the falling massive particle
 reads $T_{U}=\f{1}{2\pi\ap} $ at $t=0$,
we can rewrite this relation (\ref{ienergy}) into
\be
T_{U}\cdot \Delta S^{max}\sim E,
\ee
which looks like the thermodynamical first law. We can speculate that this is the thermodynamical
relation which is understood as that for a Rindler observer, whose acceleration is $\ap^{-1}$.
This interpretation makes sense near $t=0$ where the energy from the viewpoint of the Rindler observer coincides with $E$.

\subsection{Entanglement Renormalization and Origin of Gravitational Force}

A helpful framework to study the connection between quantum entanglement of a given quantum many-body system and its holographic geometry is the entanglement renormalization
(or called multi-scale entanglement renormalization ansatz, MERA) \cite{MERA}.
Indeed, it has been conjectured in \cite{Swingle} that the geometry described by a tensor network of a quantum critical system in MERA describes the AdS spacetime dual to this critical point. In \cite{NRT}, this correspondence has been studied in the continuum limit (i.e. field theory limit), called cMERA \cite{cMERA} and a candidate of holographic metric in terms of purely field theoretic data has been proposed. Refer also to \cite{MS} for recent developments in this subject.

Motivated by this it is intriguing to see how we can describe the time evolution of local quenches from the viewpoint of the entanglement renormalization. A general framework to
describe a quantum state graphically is called tensor networks (see e.g. reviews \cite{TNG}) and has been recently used to prepare a candidate of approximate ground states based on the variational principle,
optimized by a number of parameters.

\begin{figure}[ttt]
   \begin{center}
     \includegraphics[height=6cm]{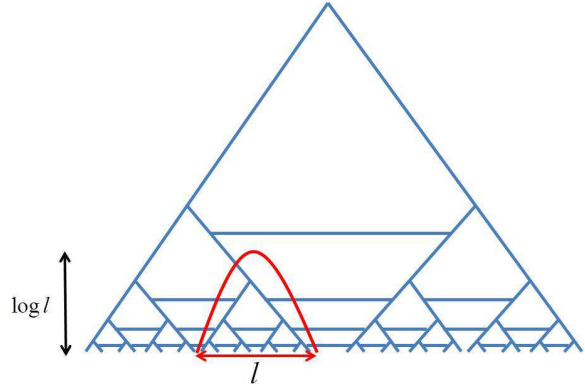}
   \end{center}
   \caption{The tensor network of MERA for a ground state of a CFT.
   The coarse-graining procedure is done from the bottom (UV) to the top (IR).
   The horizontal ladders represent the disentanglers,
   while the tree structure does the coarse-graining procedure,
   combining two sites into one. We also show the estimation of entanglement
   entropy when the subsystem $A$ is an interval with length $l$ by
   counting the number of bonds.
    }\label{fig:MERA}
\end{figure}

The tensor network for a CFT is called MERA and the network for its ground state is described in Fig.\ref{fig:MERA}. It describes the coarse-graining procedure of a given quantum system such as spin chains as we goes from the bottom (UV) to the top (IR), by combining two sites (or spins) into one, which looks like a tree in Fig.\ref{fig:MERA}.
To realize the entanglement structure of CFTs, we need to add so called disentanglers which are unitary transformations between two spins and which are written as the horizontal ladders in Fig.\ref{fig:MERA}.

Let us define the number of disentanglers such that each of them
carries the entanglement entropy $\log 2$. Then we find that the
number of disentanglers in each bond of the MERA network
 is given by $\f{c}{6}$. Indeed, this reproduces the correct entanglement entropy for ground states in CFTs given by the well-known formula (\ref{frgr}) as explained in Fig.\ref{fig:MERA}. Notice that we can estimate the entanglement entropy from the MERA
 diagram by encircling the subsystem $A$ and counting the number of intersecting bonds
 (disentanglers). Though there are many choices of encircling curve, we can optimize the result by choosing the one with minimal number of intersecting bonds. This minimizing procedure nicely matches with the HEE formula (\ref{HEE}) as pointed out in \cite{Swingle}.

 By comparing this estimation of the entanglement entropy in CFTs with (\ref{Ncft}),
 we can speculate that the number of disentanglers in a bond is roughly approximated by $l^2\cdot
 \ov{n(l,\xi,t)}$ in more general tensor networks which do not correspond to ground states of CFTs. Here the averaging symbol means an average like $\sim \f{1}{l^2}\int^l_0 dl \int^{\xi+l/2}_{\xi-l/2}d\xi$. The conservation law (\ref{necons}) tells us that the total number of disentangles is also conserved. Moreover, we can roughly speculate the relation between the entanglement density and the holographic metric in the extra direction $z$: $\s{g_{zz}(z,x,t)}\propto l\cdot\ov{n(l,\xi,t)}$ with the identification $l=z$ and $\xi=x$. It will be an interesting future problem to work out these relations more precisely.

\begin{figure}[ttt]
   \begin{center}
     \includegraphics[height=5cm]{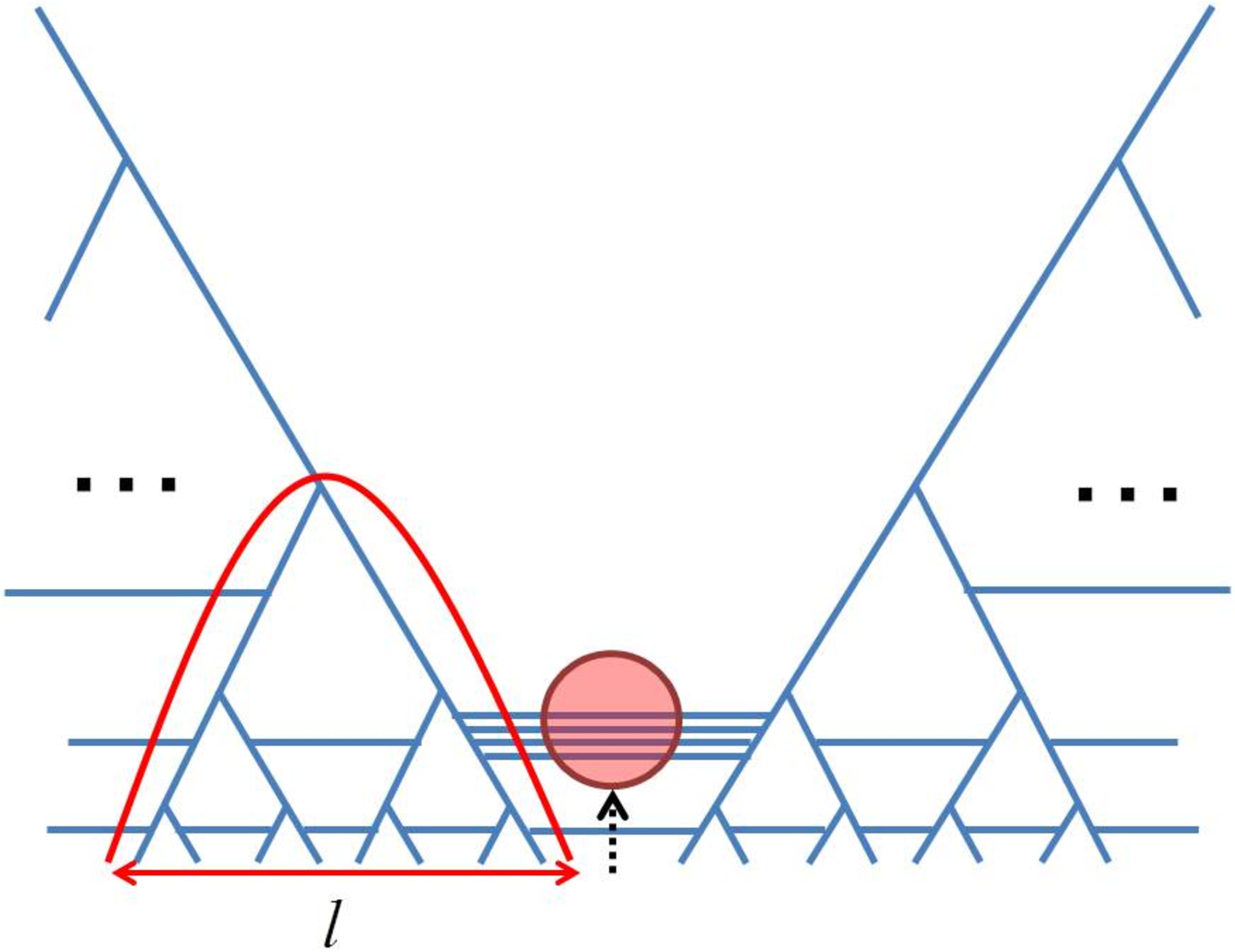}
     \hspace{5mm}
       \includegraphics[height=5cm]{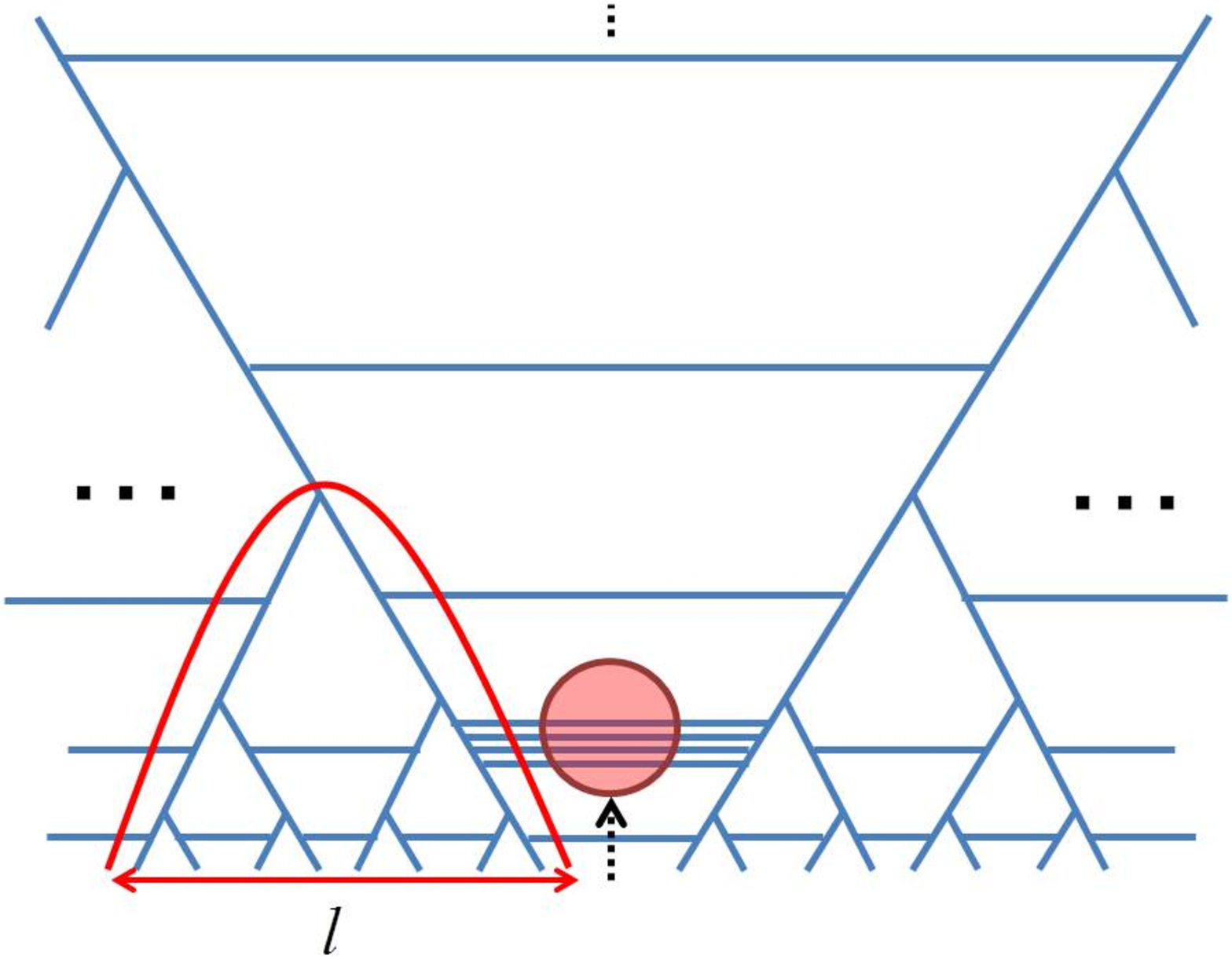}
   \end{center}
   \caption{The tensor networks of MERA after local quenches in a CFT. The left picture
   corresponds to the local quench triggered by joining two CFTs on semi-infinite lines.
   The right one describes our holographic local quench induced by local excitations in the CFT.
    }\label{fig:MERAQ}
\end{figure}

Now we would like to consider how to describe local quenches by using tensor networks. We consider both our holographic local quench and the original local quench in \cite{CaCaL}. We argue that they are described by the left and right pictures of Fig.\ref{fig:MERAQ}, respectively. In our model, there are disentanglers even in IR regions because we start with the ground state of a CFT on an infinite line.  On the other hand, in the local quench induced by the joining two semi-infinite lines, there should be no entanglement between the two in the IR region. As in (\ref{frgr}) and (\ref{eegr}), we can confirm that the behavior of $S_A$ when $t>>l$ is the same as that for the ground state. Thus the MERA structures in the UV region are also the same as that for the ground state in both cases. Finally, the massive falling particle corresponds to the insertion of many disentanglers.
This description is consistent with the behaviors (\ref{btlo}) and (\ref{btloo}) at late time.

Finally we would like to ask what is the holographic origin of gravitational force.
As we have seen, the propagation of entangled pairs is dual to the falling particle in AdS. This tells us that the evolution of the range of entanglement is dual to the gravitational force in AdS space. Therefore we can argue that the gravitational force is dual to a kind of decoherence which breaks the short range entanglement and
tries to expand the range of entanglement. In MERA, this is described by the motion of the dense lump of disentanglers toward upper direction as depicted in Fig.\ref{fig:MERAQ}.

\section{Conclusions and Discussions}

In the first half of this paper, we proposed a holographic model of
local quench and study its property especially focusing on the time
evolution of entanglement entropy. In our model, the local quench is
simply described by a free falling massive particle in an AdS space.
We performed a perturbative analysis of its back-reacted geometry
and computed the holographic entanglement entropy (HEE) in any
dimension as well as the holographic energy stress tensor. Moreover,
we presented exact calculations of HEE for local quenches in two
dimensions. We leave exact calculations in higher dimensional cases
for a future problem. Since in our model we considered a massive
particle without any charge, it will be another interesting future
problem to extend our results to charged massive particles.

In most of earlier works on local quenches, local excitations are
generated by joining two semi-infinite lines (or spin chains). On
the other hands, our holographic model corresponds to the setup
where local excitations are induced in a CFT on an infinite line and
does not involve any joining procedure. Therefore we found that
details of the time-evolution of entanglement entropy are
quantitatively different between them, while we find that the
qualitative behavior agrees with each other. For example, the
logarithmic time evolution $\sim \log t$ is common to both of them
for a large subsystem. However, its coefficients differ by the
factor two between them and we give an simple explanation from the
viewpoint of entanglement renormalization. This is because there
exists long range entanglement even just after the quench. Even
though we were able to reproduce the basic properties of local
quenches in our model, it is an intriguing future problem to
construct a clean gravity dual without any long range entanglement.

In order to study the time evolution of quantum entanglement more
clearly, we introduced a new quantity called the entanglement
density for two dimensional field theories. This measures the
density of entangled pairs between given two points. We showed that
the strong subadditivity guarantees that this quantity is positive
as should be so if we want to understand it as a physical density.
Moreover, we found a simple relation between the entanglement
density and the energy density in the small subsystem limit.

 It will be interesting to generalize this quantity so that
we can incorporate the effect of three-body or higher entanglement.
At least it is already clear that $N$-body entanglement is related
to a $N$-th derivative of the entanglement entropy for a suitable
subsystem. At the same time, it will be an important problem to
extend the entanglement density into higher dimensional counterparts
and study various holographic models. A special example of higher
dimensional entanglement density is obtained by replacing the
interval subsystems with strip ones.

In the final part of this paper, we calculated the amount of quantum
information carried by a massive object (localized excitations) in
the dual CFT. We find a rather simple rule that this amount of
information is given by the energy $E$ of the object times it linear
size $\ap$. This is expected to be a prediction which is true for
large $N$ strongly coupled gauge theories. It will very intriguing
to see how this behavior is universal in more general cases.

\section*{Acknowledgements}
TT would like to thank E. Tonni for earlier stimulating
collaborations on local quenches and valuable remarks on the draft
of this paper. We are also grateful to M. Fujita, M. Hotta, V. Hubeny, L-Y.
Hung, H. Matsueda, M. Rangamani and T. Ugajin for useful discussions, and
especially to R. Myers for reading the draft of this paper and
giving us useful comments. MN would like to thank the Perimeter Institute for giving the opportunities to discuss with researchers.
TT is supported by JSPS Grant-in-Aid for Challenging
Exploratory Research No.24654057.  TT is also
supported by World Premier International
Research Center Initiative (WPI Initiative) from the Japan Ministry
of Education, Culture, Sports, Science and Technology (MEXT).

\appendix

\section{Perturbative Construction of Gravity Duals}

In this appendix, we will show a calculation of the back-reaction for a falling massive particle in AdS$_4$ by solving (\ref{EMM}) and (\ref{emtsnr}) directly.

\subsection{Perturbation Theory}

We focus on $d=3$ i.e. the AdS$_4$ case and define $x=x_1$ and $y=x_2$. We will set $R=1$ just for simplicity.
We perform the Fourier transformation with respect to $x,y$ and $t$.
Using the symmetry $x\to -x$ and $y\to -y$, we can write down the following ansatz of the perturbed metric:
\ba
&& \delta g_{tt}=h(z)e^{i(k_x\cdot x+k_y\cdot y-\omega t)},\ \
\delta g_{tx}=\f{k_x}{\omega}\cdot w(z)e^{i(k_x\cdot x+k_y\cdot y-\omega t)},\ \
\delta g_{ty}=\f{k_y}{\omega}\cdot w(z)e^{i(k_x\cdot x+k_y\cdot y-\omega t)},\no
&& \delta g_{xx}=\left(k_x^2f(z)+g(z)\right)e^{i(k_x\cdot x+k_y\cdot y-\omega t)}, \ \
\delta g_{xy}=k_xk_y f(z)e^{i(k_x\cdot x+k_y\cdot y-\omega t)},\no
&& \delta g_{yy}=\left(k_y^2f(z)+g(z)\right) e^{i(k_x\cdot x+k_y\cdot y-\omega t)}.
\ea

The energy stress tensor for the falling particle described by $z(t)=\s{t^2+\ap^2}$ after the Fourier transformation
\be
T^{\mu\nu}(k_x,k_y,\omega,z)= \f{1}{(2\pi)^3}\int dxdydt~ e^{-i(k_x\cdot x+k_y\cdot y-\omega t)}\cdot T^{\mu\nu}(x,y,t,z).
\ee

is given by
\ba
&& T^{zz}=2\hat{m}\cdot \f{z^5\s{z^2-\ap^2}}{\ap}   \cdot \cos(\omega \s{z^2-\ap^2}),\no
&& T^{zt}=2i\hat{m}\cdot \f{z^6}{\ap}   \cdot \sin(\omega \s{z^2-\ap^2}),\no
&& T^{tt}=2\hat{m}\cdot \f{z^7}{\ap\s{z^2-\ap^2}}   \cdot \cos(\omega \s{z^2-\ap^2}),\no
\ea
and other components are vanishing. We employed the identity
\be
\delta (z-\s{t^2+\ap^2})=\f{z}{\s{z^2-\ap^2}}
\left(\delta(t-\s{z^2-\ap^2})+\delta(t+\s{z^2-\ap^2})\right),
\ee
where we assumed $z\geq 0$. We also defined `normalized mass'
\be
\hat{m}\equiv \f{8\pi G_Nm}{(2\pi)^3}.
\ee

It is straightforward to check that they satisfy the conservation law $\nabla_\mu T^{\mu\nu}=0$.

\subsection{General Solutions to Einstein Equation}

By evaluating the left-hand side of (\ref{EMM}), in the end we find that all the solutions can be constructed as follows:
First we define $u(z)$ such that
\ba
&& w(z)=A(1-\omega)+\f{\omega^2}{2}u(z)-\f{u'(z)}{z}+\f{u''(z)}{2},\no
&& f(z)=-\f{A}{\omega}-\f{u(z)}{2}-\f{u'(z)}{z\omega^2}+\f{u''(z)}{2\omega^2},\no
&& g(z)=\f{u'(z)}{z}+\f{4u'(z)}{z^3\omega^2}-\f{u''(z)}{2}-\f{4u''(z)}{z^2\omega^2}
+\f{2u^{(3)}(z)}{z\omega^2}-\f{u^{(4)}(z)}{2\omega^2},\no
&& h(z)=A\omega-\f{\omega^2}{2}u(z)+\f{2u'(z)}{z}+\f{4u'(z)}{\omega^2z^3}-u''(z)
-\f{4u''(z)}{\omega^2z^2}+\f{2u^{(3)}}{z\omega^2}-\f{u^{(4)}}{2\omega^2}. \label{sss}
\ea
Moreover, we define $y(z)$ by
\be
y(z)=\f{u'(z)}{z}.
\ee
Then we can show that (\ref{sss}) is a solution to (\ref{EMM}) if the following equation is satisfied
\be
y^{(3)}(z)-(k_x^2+k_y^2-\omega^2)y'(z)
-\f{4\hat{m}}{\omega}\left[\sin\left(\omega\s{z^2-\ap^2}\right)
-\omega\s{z^2-\ap^2}\cos\left(\omega\s{z^2-\ap^2}\right)\right]=0. \label{eomy}
\ee

It is clear that there are five integration constants. They are dual to
the background metric perturbation (four parameters) and the boundary energy stress tensor (one parameter).

\subsection{Special Solutions: $\ap\to 0$}

We can solve (\ref{eomy}) in the particular case $\ap\to 0$ where $z(t)\simeq |t|$.
We impose the boundary conditions such that there are no non-normalize deformations
(i.e. the background metric is not perturbed) and that the metric perturbation does not
blow up in the IR limit $z\to \infty$. This uniquely fixed the form of $u(z)$ as follows
\ba
 u(z)&=&-\f{4\hat{m}}{\ap\omega^4(k_x^2+k_y^2-\omega^2)^2}\Biggl(4k_x^2+4k_y^2-6\omega^2
 +z^2\omega^2(k_x^2+k_y^2-\omega^2) \no
&&+ \f{(k_x^2+k_y^2-\omega^2)^2}{(k_x^2+k_y^2)^2}
\left(\left(-4k_x^2-4k_y^2-2\omega^2+(k_x^2+k_y^2)z^2\omega^2\right)\cos(z\omega)
-2z\omega(2k_x^2+2k_y^2+\omega^2)\sin(z\omega)\right) \no
 &&+\f{1}{(k_x^2+k_y^2)^2}\cdot 2e^{-z\s{k_x^2+k_y^2-\omega^2}}
 \omega^6\left(1+z\s{k_x^2+k_y^2-\omega^2}\right)\Biggr).
\ea
We also have the condition $A=0$. These completely fix $f(z),g(z),h(z)$ and $w(z)$.

Near the AdS boundary $z\to 0$ they behave
\ba
&& w(z)\simeq -\f{4\hat{m}\omega^2}{3\ap(k_x^2+k_y^2)\s{k_x^2+k_y^2-\omega^2}}\cdot z^3,\no
&& f(z)\simeq -\f{4\hat{m}(k_x^2+k_y^2-2\omega^2)}{3\ap(k_x^2+k_y^2)^2\s{k_x^2+k_y^2-\omega^2}}\cdot z^3,\no
&& g(z)\simeq \f{4\hat{m}(k_x^2+k_y^2-\omega^2)}{3\ap(k_x^2+k_y^2)\s{k_x^2+k_y^2-\omega^2}}\cdot z^3,\no
&& h(z)\simeq \f{4\hat{m}}{3\s{k_x^2+k_y^2-\omega^2}}\cdot z^3.
\ea

We can read off the tensor $t_{ab}$ defined by (\ref{emh}) as follows:
\ba
&& t_{tt}=\f{4\hat{m}}{3\ap\s{k_x^2+k_y^2-\omega^2}}, \no
&& t_{tx}=-\f{4\hat{m}k_x\omega}{3\ap(k_x^2+k_y^2)\s{k_x^2+k_y^2-\omega^2}}, \no
&& t_{ty}=-\f{4\hat{m}k_y\omega}{3\ap(k_x^2+k_y^2)\s{k_x^2+k_y^2-\omega^2}}, \no
&& t_{xx}=\f{4\hat{m}(k_y^4+k_x^2k_y^2-k_y^2\omega^2+\omega^2k_x^2)}
{3\ap(k_x^2+k_y^2)^2\s{k_x^2+k_y^2-\omega^2}}, \no
&& t_{xy}=-\f{4\hat{m}k_xk_y(k_x^2+k_y^2-2\omega^2)}
{3\ap(k_x^2+k_y^2)^2\s{k_x^2+k_y^2-\omega^2}}, \no
&& t_{yy}=\f{4\hat{m}(k_x^4+k_x^2k_y^2-k_x^2\omega^2+\omega^2k_y^2)}
{3\ap(k_x^2+k_y^2)^2\s{k_x^2+k_y^2-\omega^2}}. \label{empt}
\ea

The physical holographic stress tensor $T_{ab}$ is calculated from
(\ref{emh}) via the formula (\ref{emht}). In the end we will find that (\ref{empt})
reproduces the result (\ref{EMthree}) at $\ap=0$.
For example,  if we take the $\ap\to 0$ limit of (\ref{EMthree}), we can show
\be
T_{tt}(t,\rho)\to \f{M}{2\pi G_N R\ap}\delta(\rho^2-t^2),
\ee
by using
\be
\lim_{\ap\to 0} \f{\ap^4}{(x^2+\ap^2r^2)^{\f{5}{2}}}=\f{4}{3r^4}\delta(x).
\ee
By performing the Fourier transformation, we find
\be
T_{tt}(\omega,k_x,k_y)\to \f{M}{8\pi^3G_N R\ap} (k_x^2+k_y^2-\omega^2)^{-1/2}.
\ee
This precisely agrees with (\ref{empt}).


\begin{thebibliography}{99}

\baselineskip=8pt



\bibitem{Ereview}
  J.~Eisert, M.~Cramer and M.~B.~Plenio,
  ``Area laws for the entanglement entropy - a review,''
  Rev.\ Mod.\ Phys.\  {\bf 82} (2010) 277  [arXiv:0808.3773 [quant-ph]].

\bibitem{CCreview}
 P.~Calabrese and J.~Cardy,
 ``Entanglement entropy and conformal field theory,''
 J.\ Phys.\ A A {\bf 42} (2009) 504005  [arXiv:0905.4013 [cond-mat.stat-mech]].

\bibitem{CHreview}
  H.~Casini and M.~Huerta,
  ``Entanglement entropy in free quantum field theory,''
  J.\ Phys.\ A A {\bf 42} (2009) 504007  [arXiv:0905.2562 [hep-th]].

%

\bibitem{Lreview}
  J.~I.~Latorre, A.~Riera,
  ``A short review on entanglement in quantum spin systems,''
  J.\ Phys.\ A {\bf 42} (2009) 504002  [arXiv:0906.1499 [hep-th]].



\bibitem{HEEreview}
 T.~Nishioka, S.~Ryu and T.~Takayanagi,
 ``Holographic Entanglement Entropy: An Overview,''
  J.\ Phys.\ A  {\bf 42} (2009) 504008;
 T.~Takayanagi,
  ``Entanglement Entropy from a Holographic Viewpoint,''
  Class.\ Quant.\ Grav.\  {\bf 29} (2012) 153001  [arXiv:1204.2450 [gr-qc]].



\bibitem{CaCaG}
P.~Calabrese and J.~L.~Cardy,
  ``Evolution of Entanglement Entropy in One-Dimensional Systems,''
  J.\ Stat.\ Mech.\  {\bf 04} (2005) P04010, cond-mat/0503393;
 P.~Calabrese and J.~L.~Cardy, ``Time-dependence of correlation functions following a quantum quench,''
  Phys.Rev.Lett.{\bf 96}(2006)136801;
P.~Calabrese and J.~L.~Cardy,  ``Quantum Quenches in Extended Systems,''
  J.\ Stat.\ Mech.\  {\bf 06} (2007) P06008, arXiv:0704.1880;
      S.~Sotiriadis and J.~Cardy,
  ``Quantum quench in interacting field theory: A Self-consistent approximation,''  Phys.\ Rev.\ B {\bf 81} (2010) 134305  [arXiv:1002.0167 [quant-ph]].  



\bibitem{Eisler}
V.~Eisler, I.~Peschel,
``Evolution of entanglement after a local quench,''
 J.\ Stat.\ Mech.\ (2007) P06005, cond-mat/0703379;

 V.~Eisler, D.~Karevski, T.~Platini, I.~Peschel,
 ``Entanglement evolution after connecting finite to infinite quantum chains,''
  J.\ Stat.\ Mech.\ (2008) P01023, arXiv:0711.0289.



\bibitem{CaCaL}
P.~Calabrese and J.~L.~Cardy,
  ``Entanglement and correlation functions following a local quench: a conformal field theory approach,''
  J.\ Stat.\ Mech.\  {\bf 10} (2007) P10004, arXiv:0708.3750.


\bibitem{Maldacena}
  J.~M.~Maldacena,
  ``The large N limit of superconformal field theories and supergravity,''
  Adv.\ Theor.\ Math.\ Phys.\  {\bf 2} (1998) 231
  [Int.\ J.\ Theor.\ Phys.\  {\bf 38} (1999) 1113]
  [arXiv:hep-th/9711200];



\bibitem{GKPW}
S.~S.~Gubser, I.~R.~Klebanov and A.~M.~Polyakov,
  ``Gauge theory correlators from non-critical string theory,''
  Phys.\ Lett.\ B {\bf 428}, 105 (1998);
E.~Witten,
  ``Anti-de Sitter space and holography,''
  Adv.\ Theor.\ Math.\ Phys.\  {\bf 2}, 253 (1998).



\bibitem{AdSreview}
  O.~Aharony, S.~S.~Gubser, J.~M.~Maldacena, H.~Ooguri and Y.~Oz,
  ``Large N field theories, string theory and gravity,''  Phys.\ Rept.\  {\bf 323} (2000) 183  [hep-th/9905111].  




\bibitem{QuenchHol}
  S.~Bhattacharyya and S.~Minwalla,
  ``Weak Field Black Hole Formation in Asymptotically AdS Spacetimes,''  JHEP {\bf 0909} (2009) 034  [arXiv:0904.0464 [hep-th]];  
S.~R.~Das, T.~Nishioka and T.~Takayanagi,
  ``Probe Branes, Time-dependent Couplings and Thermalization in AdS/CFT,''  JHEP {\bf 1007} (2010) 071  [arXiv:1005.3348 [hep-th]];  
 H.~Ebrahim and M.~Headrick,
  ``Instantaneous Thermalization in Holographic Plasmas,''  arXiv:1010.5443 [hep-th];  
 D.~Garfinkle and L.~A.~Pando Zayas,
  ``Rapid Thermalization in Field Theory from Gravitational Collapse,''  Phys.\ Rev.\ D {\bf 84} (2011) 066006  [arXiv:1106.2339 [hep-th]];  
   P.~Basu and S.~R.~Das,
  ``Quantum Quench across a Holographic Critical Point,''  JHEP {\bf 1201} (2012) 103  [arXiv:1109.3909 [hep-th]];  
S.~R.~Das,
  ``Holographic Quantum Quench,''  J.\ Phys.\ Conf.\ Ser.\  {\bf 343} (2012) 012027  [arXiv:1111.7275 [hep-th]];  
  E.~Caceres and A.~Kundu,
  ``Holographic Thermalization with Chemical Potential,''
  JHEP {\bf 1209} (2012) 055
  [arXiv:1205.2354 [hep-th]];
A.~Buchel, L.~Lehner and R.~C.~Myers,
  ``Thermal quenches in N=2* plasmas,''  JHEP {\bf 1208} (2012) 049  [arXiv:1206.6785 [hep-th]];
 M.~J.~Bhaseen, J.~P.~Gauntlett, B.~D.~Simons, J.~Sonner and T.~Wiseman,
  ``Holographic Superfluids and the Dynamics of Symmetry Breaking,''  arXiv:1207.4194 [hep-th].  
   P.~Basu, D.~Das, S.~R.~Das and T.~Nishioka,
  ``Quantum Quench Across a Zero Temperature Holographic Superfluid Transition,''  arXiv:1211.7076 [hep-th].  
 X.~Gao, A.~M.~Garcia-Garcia, H.~B.~Zeng and H.~-Q.~Zhang,
  ``Lack of thermalization in holographic superconductivity,''  arXiv:1212.1049 [hep-th].  
 W.~H.~Baron, D.~Galante and M.~Schvellinger,
  ``Dynamics of holographic thermalization,''  arXiv:1212.5234 [hep-th];  
 A.~Buchel, L.~Lehner, R.~C.~Myers and A.~van Niekerk,
  ``Quantum quenches of holographic plasmas,''  arXiv:1302.2924 [hep-th].  




\bibitem{QuenchHEE}
 J.~Abajo-Arrastia, J.~Aparicio and E.~Lopez,
  ``Holographic Evolution of Entanglement Entropy,''  JHEP {\bf 1011} (2010) 149  [arXiv:1006.4090 [hep-th]];  
  T.~Albash and C.~V.~Johnson,
  ``Evolution of Holographic Entanglement Entropy after Thermal and Electromagnetic Quenches,''  New J.\ Phys.\  {\bf 13} (2011) 045017  [arXiv:1008.3027 [hep-th]]; 
   V.~Keranen, E.~Keski-Vakkuri and L.~Thorlacius,
  ``Thermalization and entanglement following a non-relativistic holographic quench,''  Phys.\ Rev.\ D {\bf 85} (2012) 026005  [arXiv:1110.5035 [hep-th]];  
    D.~Galante and M.~Schvellinger,
  ``Thermalization with a chemical potential from AdS spaces,''  JHEP {\bf 1207} (2012) 096  [arXiv:1205.1548 [hep-th]];  
 A.~Bernamonti, N.~Copland, B.~Craps and F.~Galli,
  ``Holographic thermalization of mutual and tripartite information in 2d CFTs,''  arXiv:1212.0848 [hep-th].  



\bibitem{TaUg}
T.~Takayanagi and T.~Ugajin,
  ``Measuring Black Hole Formations by Entanglement Entropy via Coarse-Graining,''  JHEP {\bf 1011} (2010) 054  [arXiv:1008.3439 [hep-th]].  


\bibitem{Roberts:2012aq}
  M.~M.~Roberts,
  ``Time evolution of entanglement entropy from a pulse,''  arXiv:1204.1982 [hep-th].




\bibitem{RT}
  S.~Ryu and T.~Takayanagi,
  ``Holographic derivation of entanglement entropy from AdS/CFT,''
  Phys.\ Rev.\ Lett.\  {\bf 96} (2006) 181602;
 ``Aspects of holographic entanglement entropy,''
  JHEP {\bf 0608} (2006) 045.


\bibitem{HRT} V.~E.~Hubeny, M.~Rangamani and T.~Takayanagi, ``A Covariant
holographic entanglement entropy proposal,'' JHEP {\bf 0707} (2007)
062  [arXiv:0705.0016 [hep-th]].




\bibitem{Bla}
  V.~Balasubramanian, A.~Bernamonti, J.~de Boer, N.~Copland, B.~Craps, E.~Keski-Vakkuri, B.~Muller and A.~Schafer {\it et al.},
  ``Thermalization of Strongly Coupled Field Theories,''  Phys.\ Rev.\ Lett.\  {\bf 106} (2011) 191601  [arXiv:1012.4753 [hep-th]].  
``Holographic Thermalization,''  Phys.\ Rev.\ D {\bf 84} (2011) 026010  [arXiv:1103.2683 [hep-th]].  

\bibitem{TSSA}
 A.~Allais and E.~Tonni,
  ``Holographic evolution of the mutual information,''  JHEP {\bf 1201} (2012) 102  [arXiv:1110.1607 [hep-th]];  
 R.~Callan, J.~-Y.~He and M.~Headrick,
  ``Strong subadditivity and the covariant holographic entanglement entropy formula,''  JHEP {\bf 1206} (2012) 081  [arXiv:1204.2309 [hep-th]].  
  A.~C.~Wall,
  ``Maximin Surfaces, and the Strong Subadditivity of the Covariant Holographic Entanglement Entropy,''  arXiv:1211.3494 [hep-th].  

\bibitem{NTT}
 K.~Narayan, T.~Takayanagi and S.~P.~Trivedi,
  ``AdS plane waves and entanglement entropy,''  arXiv:1212.4328 [hep-th].  

\bibitem{HoIt}
  G.~T.~Horowitz and N.~Itzhaki,
  ``Black holes, shock waves, and causality in the AdS / CFT correspondence,''  JHEP {\bf 9902} (1999) 010  [hep-th/9901012].  


\bibitem{DKK}
  U.~H.~Danielsson, E.~Keski-Vakkuri and M.~Kruczenski,
  ``Vacua, propagators, and holographic probes in AdS / CFT,''  JHEP {\bf 9901} (1999) 002  [hep-th/9812007].  



\bibitem{FGMP}
  J.~J.~Friess, S.~S.~Gubser, G.~Michalogiorgakis and S.~S.~Pufu,
  ``Expanding plasmas and quasinormal modes of anti-de Sitter black holes,''  JHEP {\bf 0704} (2007) 080  [hep-th/0611005].  

\bibitem{Ran}
  P.~Figueras, V.~E.~Hubeny, M.~Rangamani and S.~F.~Ross,
  ``Dynamical black holes and expanding plasmas,''  JHEP {\bf 0904} (2009) 137  [arXiv:0902.4696 [hep-th]].  



\bibitem{Witten}
  E.~Witten,
  ``Anti-de Sitter space, thermal phase transition, and confinement in gauge theories,''  Adv.\ Theor.\ Math.\ Phys.\  {\bf 2} (1998) 505  [hep-th/9803131].  


\bibitem{EMtensor}
V.~Balasubramanian and P.~Kraus,
  ``A Stress tensor for Anti-de Sitter gravity,''
  Commun.\ Math.\ Phys.\  {\bf 208} (1999) 413;
 S.~de Haro, S.~N.~Solodukhin and K.~Skenderis,
  ``Holographic reconstruction of space-time and renormalization in the AdS /
  CFT correspondence,''
  Commun.\ Math.\ Phys.\  {\bf 217} (2001) 595.

\bibitem{CHM}
 H.~Casini, M.~Huerta and R.~C.~Myers,
  ``Towards a derivation of holographic entanglement entropy,''  JHEP {\bf 1105} (2011) 036  [arXiv:1102.0440 [hep-th]].  

\bibitem{Area}
  L.~Bombelli, R.~K.~Koul, J.~Lee and R.~D.~Sorkin,
  ``A Quantum Source of Entropy for Black Holes,''
  Phys.\ Rev.\ D {\bf 34} (1986) 373;
   M.~Srednicki,
   ``Entropy and area,''
   Phys.\ Rev.\ Lett.\  {\bf 71} (1993) 666  [hep-th/9303048].


\bibitem{BNTU}
  J.~Bhattacharya, M.~Nozaki, T.~Takayanagi and T.~Ugajin,
  ``Thermodynamical Property of Entanglement Entropy for Excited States,''  arXiv:1212.1164 [hep-th].  


\bibitem{BrHe}
 J.~D.~Brown and M.~Henneaux,
  ``Central Charges in the Canonical Realization of
  Asymptotic Symmetries: An Example from Three-Dimensional Gravity,''
  Commun.\ Math.\ Phys.\  {\bf 104} (1986) 207.


\bibitem{HLW}
C.~Holzhey, F.~Larsen and F.~Wilczek,
  ``Geometric and renormalized entropy in conformal field theory,''
  Nucl.\ Phys.\ B {\bf 424}, 443 (1994)
  [arXiv:hep-th/9403108].

\bibitem{CaCa}
  P.~Calabrese and J.~Cardy,
  ``Entanglement entropy and quantum field theory,''
  J.\ Stat.\ Mech.\  {\bf 0406}, P002 (2004)
  [arXiv:hep-th/0405152].





\bibitem{LR}
  E.~H.~Lieb and M.~B.~Ruskai,
  ``Proof of the strong subadditivity of quantum-mechanical entropy,''  J.\ Math.\ Phys.\  {\bf 14} (1973) 1938.


\bibitem{CHS}
H.~Casini and M.~Huerta, ``A Finite entanglement entropy and the
c-theorem,''  Phys.\ Lett.\ B {\bf 600} (2004) 142
[hep-th/0405111].

\bibitem{HT}
  M.~Headrick and T.~Takayanagi,
  ``A Holographic proof of the strong subadditivity of entanglement entropy,''  Phys.\ Rev.\ D {\bf 76} (2007) 106013  [arXiv:0704.3719 [hep-th]].


\bibitem{FTT}
  M.~Fujita, T.~Takayanagi and E.~Tonni,
  ``Aspects of AdS/BCFT,''  JHEP {\bf 1111} (2011) 043  [arXiv:1108.5152 [hep-th]].  



\bibitem{Ta}
  T.~Takayanagi,
  ``Holographic Dual of BCFT,''  Phys.\ Rev.\ Lett.\  {\bf 107} (2011) 101602  [arXiv:1105.5165 [hep-th]].  



\bibitem{Hage}
  B.~Sundborg,
  ``The Hagedorn transition, deconfinement and N=4 SYM theory,''  Nucl.\ Phys.\ B {\bf 573} (2000) 349  [hep-th/9908001]; 

  O.~Aharony, J.~Marsano, S.~Minwalla, K.~Papadodimas and M.~Van Raamsdonk,
  ``The Hagedorn - deconfinement phase transition in weakly coupled large N gauge theories,''  Adv.\ Theor.\ Math.\ Phys.\  {\bf 8} (2004) 603  [hep-th/0310285].  


\bibitem{MERA}
G.~Vidal,``Entanglement renormalization,''
Phys. Rev. Lett. {\bf 99}, 220405 (2007) , arXiv:cond-mat/0512165;
  ``Entanglement renormalization: an introduction,''
arXiv:0912.1651;
G.~Evenbly and G.~Vidal,
``Quantum Criticality with the Multi-scale Entanglement Renormalization Ansatz,'' arXiv:1109.5334.



\bibitem{Swingle}
  B.~Swingle,
  ``Entanglement Renormalization and Holography,''
  Phys.Rev.D {\bf 86} (2012)065007,
  arXiv:0905.1317 [cond-mat.str-el];  
 ``Constructing holographic spacetimes using entanglement renormalization,''  arXiv:1209.3304 [hep-th].  





\bibitem{NRT}
  M.~Nozaki, S.~Ryu and T.~Takayanagi,
  ``Holographic Geometry of Entanglement Renormalization in Quantum Field Theories,''  arXiv:1208.3469 [hep-th].  





\bibitem{cMERA}
  J.~Haegeman, T.~J.~Osborne, H.~Verschelde and F.~Verstraete,
  ``Entanglement renormalization for quantum fields,''  arXiv:1102.5524 [hep-th].



\bibitem{MS}
J.~Molina-Vilaplana and P.~Sodano,
``Holographic View on Quantum Correlations and Mutual Information between Disjoint Blocks of a Quantum Critical System,''  JHEP {\bf 1110} (2011) 011  [arXiv:1108.1277 [quant-ph]];  
  V.~Balasubramanian, M.~B.~McDermott and M.~Van Raamsdonk,
  ``Momentum-space entanglement and renormalization in quantum field theory,''  arXiv:1108.3568 [hep-th];  
H.~Matsueda,``Scaling of entanglement entropy and hyperbolic geometry,''
 arXiv:1112.5566 [cond-mat.stat-mech]; 
  M.~Ishihara, F.~-L.~Lin and B.~Ning,
  ``Refined Holographic Entanglement Entropy for the AdS Solitons and AdS black Holes,''  arXiv:1203.6153 [hep-th];  
H.~Matsueda, M.~Ishihara and Y.~Hashizume,
``Tensor Network and Black Hole,''
arXiv:1208.0206 [hep-th].


\bibitem{TNG}
  J. Ignacio Cirac and Frank Verstraete,
  ``Renormalization and tensor product states in spin chains and lattices,''
J.\ Phys.\ A \textbf{42} 504004, (2009), arXiv:0910.1130;
G.~Evenbly and G.~Vidal,``Tensor network states and geometry,''
arXiv:1106.1082.





\end{thebibliography}
\end{document}